\documentclass[11pt]{article}
\pdfoutput=1
\usepackage{geometry}             
\usepackage{graphicx}
\usepackage{amssymb}
\usepackage{amsmath}
\usepackage{epstopdf}
\usepackage{comment}
\usepackage{cite}
\usepackage[font={scriptsize,singlespacing}]{caption}

\newcommand{\be}{\begin{equation}}
\newcommand{\ee}{\end{equation}}
\newcommand{\bea}{\begin{eqnarray}}
\newcommand{\eea}{\end{eqnarray}}

\usepackage{hyperref}
\usepackage{url}
\usepackage{xcolor}
\usepackage{color}
\usepackage{mathrsfs}
\usepackage{calrsfs}
\usepackage{amsfonts}
\usepackage{latexsym}
\usepackage{ragged2e}
\usepackage{epsfig}
\usepackage{textcomp}

\definecolor{vividviolet}{rgb}{0.62, 0.0, 1.0}
\definecolor{amaranth}{rgb}{0.9, 0.17, 0.31}
\definecolor{palatinateblue}{rgb}{0.15, 0.23, 0.89}
\definecolor{brightpink}{rgb}{1.0, 0.0, 0.5}
\definecolor{cornflowerblue}{rgb}{0.39, 0.58, 0.93}
\definecolor{deepcarminepink}{rgb}{0.94, 0.19, 0.22}
\definecolor{radicalred}{rgb}{1.0, 0.21, 0.37}
\hypersetup{ linktoc=all,
    colorlinks, linkcolor={palatinateblue},
    citecolor={brightpink}, urlcolor={amaranth}
}

\graphicspath{{Images/}}
\renewcommand{\d}[1]{\ensuremath{\operatorname{d}\!{#1}}}

\makeatletter
\g@addto@macro\bfseries{\boldmath}
\makeatother

\begin{document}

\thispagestyle{empty}
\begin{center}

\null \vskip-1truecm \vskip2truecm

{\Large{\bf \textsf{Hawking Evaporation of Black Holes in Massive Gravity}}}
\newline

{\textsf{Meng-Shi Hou, Hao Xu$^*$, Yen Chin Ong$^\dagger$}}\\
\vskip0.1truecm

{Center for Gravitation and Cosmology, \\College of Physical Science and Technology, Yangzhou University, \\180 Siwangting Road, Yangzhou city, Jiangsu Province 225002, China}\\
{\tt Email: haoxu\_phys@163.com$^*$, ycong@yzu.edu.cn$^\dagger$}\\

\end{center}
\vskip1truecm \centerline{\textsf{ABSTRACT}} \baselineskip=15pt

\medskip
We study the Hawking evaporation of a class of black hole solutions in dRGT massive gravity, in which the graviton mass gives rise to an effective negative cosmological constant. We found that the effective emission surface can be either proportional to the square of the effective AdS length scale, or corresponds to the square of the impact parameter of the null geodesic that falls onto the photon orbit of the black hole. Furthermore, depending on the black hole parameters, the emission surface could switch from one to another as the black hole loses mass during the evaporation process. Furthermore, the black holes can either evaporate completely or become a remnant at late time.
Our result is more generally applicable to any asymptotically anti-de Sitter-like black hole solution in any theory whose metric function has a term linear in the coordinate radius, with massive gravity being only a concrete example.

\section{Introduction to Massive Gravity}

Although general relativity has successfully described physics within the scale of the Solar System, there are still many unsolved puzzles when it is applied to a larger scale, such as the inconsistencies with the observation of galactic rotation curves and the accelerated expansion of the universe. Consequently, unknown entities, namely ``dark matter'' and ``dark energy'', were introduced to explain these anomalies. However, despite the great efforts of searching for dark matter and dark energy candidates, their true identities still remain unknown. Thus, one may wonder whether it is possible to modify the theory of gravity to explain the physics at those larger scales, while maintaining the known behaviors at the scale of the Solar System. In other words, a viable theory of modified gravity should not only explain away dark matter and/or dark energy, it should also reduce to general relativity in the regime that the latter is well-tested.

For example, one of the candidates of these modified gravity theory is conformal (Weyl) gravity \cite{Maldacena:2011mk,Anastasiou:2016jix}, which has been shown to be perturbatively renormalizable in four dimensions and can produce the effective potential consistent with the observed phenomena \cite{Stelle:1976gc,Mannheim:2009qi,Mannheim:2005bfa,Mannheim:2010ti,Mannheim:2011ds}. However, conformal (Weyl) gravity is described by a pure Weyl squared action and the field equations are fourth-orders, so it will introduce a ghost, leading to a violation of unitarity. Indeed, according to the theorem of Ostrogradsky \cite{Woodard:2015zca}, such a system is not kinematically stable.

Another candidate of modified gravity theory is massive gravity, which is an extension of general relativity by endowing graviton with a nonzero mass \cite{Fierz:1939ix,VanNieuwenhuizen:1973fi,vanDam:1970vg,Zakharov:1970cc,Vainshtein:1972sx,Boulware:1973my,deRham:2010ik,deRham:2010kj}. According to the representation theory of the Poincar\'e's group in four dimensions, any massive spin-2 state has 5 degrees of freedom, which corresponds to the helicity $0,\pm1,\pm2$ states. The correct massive gravity action should be able to describe these states. The first attempt to derive such a theory was done in 1939 by Fierz and Pauli \cite{Fierz:1939ix}. They added -- the only ghost-free and tachyon-free -- interaction terms in the linearized level of general relativity, that describe all the above 5 states. However, their theory suffered from a discontinuity in its predictions: in the massless limit the theory does not reduce to general relativity. This is known as van Dam-Veltman-Zakharov (vDVZ) discontinuity \cite{VanNieuwenhuizen:1973fi,vanDam:1970vg,Zakharov:1970cc}, the result of which is that light deflection around the Sun is off by 25\%.

The vDVZ discontinuity inspired further studies to generalize Fierz-Pauli massive gravity to nonlinear cases. Vainshtein argued that the linearized theory cannot be trusted inside some characteristic length scale, now called the ``Vainshtein radius'', and the troublesome longitudinal mode can be suppressed at measurable distances by nonlinear effects, thus making the theory compatible with current observations \cite{Vainshtein:1972sx}. However, the same nonlinear terms that made the suppression works will also generate a higher derivative term in the field equation. This, much like the conformal Weyl gravity mentioned in the beginning, could potentially lead to a ghost instability in the theory. Such ghost does in fact arise -- it is known as Boulware-Deser (BD) ghost \cite{Boulware:1973my}, which acts as the 6th degree of freedom in the theory. Though infinitely heavy on the Minkowski background, it becomes sufficiently light and propagates on locally nontrivial backgrounds.

The existence of BD ghost essentially killed off the idea of massive gravity for many decades, until it was revived again in recent years, by introducing even more nonlinear terms to exorcise the BD ghost.
It was later proved that the theory is indeed ghost-free, and furthermore the field equation is at most second order in time derivatives.
This theory, which was constructed by de Rham, Gabadadze, and Tolley (dRGT) \cite{deRham:2010ik,deRham:2010kj} (ghost-freeness was proved by Hassan and Rosen in \cite{Hassan,HassanI,HassanII}), revived the interests in massive gravity. One necessary feature of the theory is that, in order to describe gravity as arising from gravitons, there is a need for a background metric on which the gravitons propagate. This ``fiducial'' metric is fixed and must be chosen by hand. A natural choice is of course the Minkowski background, but one must remember that for each choice of the fiducial metric one is essentially dealing with a different theory, that is, ghost-freeness cannot be guaranteed in general.

Although the nonlinear terms lead to complexity in the calculations, the advantages of dRGT gravity on both theoretical and phenomenological fronts had encouraged a wide range of investigations in the literature. It must be emphasized here, however, that dRGT gravity does suffer from other problems, including violation of causality \cite{1306.5457, 1408.0561,
1410.2289, 1504.02919, 1505.03518}, and the lack of stable FLRW solution when the background is chosen to be Minkowski \cite{DAmico, 1304.0484}. For this reason, many practitioners have moved on to bimetric theory (Hassan-Rosen theory), a generalization of massive gravity in which the background metric has also becomes dynamical \cite{1109.3515}. Bimetric theory admits good cosmological solution \cite{1808.09440} and its causal structure appears to be richer and more robust \cite{1706.07806}, though more detailed studies are required to understand the mathematical structures of this complicated theory. Nevertheless, massive gravity is still being applied in the context of holography, since the graviton mass breaks diffeomorphism invariance, and therefore can be used to model dual field theory that lacks translational invariance (on in which momentum can be dissipated), without resorting to more complicated procedures, such as introducing a lattice \cite{Vegh, 1308.4970}.

In the present work we investigate a class of $(3+1)$-dimensional spherically symmetric evaporating black holes of dRGT massive gravity in anti-de Sitter (AdS) spacetime. Studies on black hole evaporation process have been a fruitful arena for theoretical physics research since the discovery of Hawking radiation. Initial investigations mainly focused on black holes in asymptotically flat spacetimes, while those of AdS spacetimes are largely overlooked. This is because massive particles emitted from the black hole will always be reflected back by the effective potential of AdS spacetime and get re-absorbed by the black hole. Massless particles can reach the boundary of AdS and get reflected back as well if we impose the natural reflective boundary condition. Thus large AdS black holes can reach thermal equilibrium with their own Hawking radiation. However, if we choose an absorbing AdS boundary condition (corresponding to coupling the field theory at the boundary with another auxiliary system that absorbs the radiation), the black hole would evaporate \cite{0804.0055, 1304.6483, 1307.1796}, and some can even evaporate completely. In \cite{1507.02682}, Page adopted the absorbing AdS boundary condition and applied the geometrical optics approximation to study the spherically symmetric AdS black hole in Einstein's gravity. One finds that the lifetime of an arbitrarily large black hole is bounded by a time of the order $\ell^{3}$, where $\ell$ is the AdS curvature radius. This is in stark contrast with the asymptotically flat case in which black hole lifetime goes as $M^3$, where $M$ is the initial mass of the black hole. Recent studies have also extended this to more complicated asymptotically AdS black holes \cite{Ong:2015fha,Xu:2017ahm,Xu:2018liy,Xu:2019krv,Xu:2020xsl}.

In this work we shall study the Hawking evaporation of the dRGT massive black hole solutions, with their thermodynamical properties given in \cite{Ghosh:2015cva}. The graviton mass generates three terms in the black hole metric, which are, respectively, an effective cosmological constant term, a linear term (linear in coordinate radius), and a ``global monopole'' term. In the massless limit the black hole solution reduces to the asymptotically \emph{flat} Schwarzschild case. The thermodynamical properties are also modified, depending on the features of these three terms. Unlike the Schwarzschild-AdS case, where there is always a Hawking-Page phase transition and the black hole can evaporate away in a finite time, the black hole thermodynamics in dRGT massive gravity is quite rich. For some values of the black hole parameters, there can be more than one horizon, and the existence of black hole remnant at late time \cite{EslamPanah:2018rob}. Thus the evolution of black holes under Hawking evaporation will also be different from the Schwarzschild-AdS case. In fact, the effective emission surface can be proportional to the square of the effective AdS length scale, or to the square of the impact parameter corresponding to the photon orbit. It is also possible that one emission surface changes to another one as the black hole losses its mass. We remark that while such a black hole solution arises in massive gravity, they can also be solutions to other modified theories of gravity. Thus, more generally, our study applies to black hole solutions in which there is a competition between a linear term and a cosmological constant term, e.g. in $f(R)$ gravity (for an example in which a linear term appears in $f(R)$ gravity black hole, see \cite{0708.1482}).

In the next section we give a brief review of the black hole solution and their interesting thermodynamics. In Sec.(3) we investigate the black hole evaporation process. In the last section we summarize the result. We adopt the Planck unit system, setting the speed of light in vacuum $c$, the gravitational constant $G$, the Planck constant $\hbar$ and the Boltzmann constant $k_B$ all equal to unity.

\section{Thermodynamics of dRGT Massive Gravity Black Holes}

In this section we give a brief review on a class of black hole solutions in dRGT massive gravity and its thermodynamics \cite{Ghosh:2015cva}. The action of dRGT massive gravity can be written as Hilbert-Einstein action with suitable nonlinear interaction terms given by
\begin{equation}
{I}_{\text{dRGT}}=\frac{1}{16\pi }\int \text{d}^{4}x\sqrt{-g}\left[ R+m^{2}\mathcal{U}%
(g,\phi ^{a})\right] ,  \label{action}
\end{equation}%
where $R$ is the Ricci scalar and $\mathcal{U}$ is the effective potential of graviton with nonzero graviton mass $m$. There are two things worth noting here. Firstly, despite appearance this theory should \emph{not} be viewed as a scalar-tensor theory. Here the so-called ``St\"{u}ckelberg scalars'' were introduced as a mean to restore the general covariance of the theory \cite{hgs,1105.3735}. Secondly, the cosmological constant is not introduced by hand in the action. Instead, an effective cosmological constant term in the metric emerged from the graviton mass, which is also the original motivation of the massive gravity to explain the accelerating Universe without resorting to dark energy (though that has proved a lot more challenging in practice). The effective potential $\mathcal{U}$ reads
\begin{equation}
\mathcal{U}\left( g,\phi ^{a}\right) =\mathcal{U}_{2}+\alpha _{3}\mathcal{U}%
_{3}+\alpha _{4}\mathcal{U}_{4},  \label{U}
\end{equation}%
where $\alpha _{3}$ and $\alpha _{4}$ are dimensionless free coefficients, and
\begin{eqnarray}
\mathcal{U}_{2} &=&\left[ \mathcal{K}\right] ^{2}-\left[ \mathcal{K}^{2}%
\right] ,  \notag \\
\mathcal{U}_{3} &=&\left[ \mathcal{K}\right] ^{3}-3\left[ \mathcal{K}\right] %
\left[ \mathcal{K}^{2}\right] +2\left[ \mathcal{K}^{3}\right] ,  \notag \\
\mathcal{U}_{4} &=&\left[ \mathcal{K}\right] ^{4}-6\left[ \mathcal{K}^{2}%
\right] \left[ \mathcal{K}\right] ^{2}+8\left[ \mathcal{K}^{3}\right] \left[
\mathcal{K}\right] +3\left[ \mathcal{K}^{2}\right] ^{2}-6\left[ \mathcal{K}^{4}\right] ,
\end{eqnarray}%
where
\begin{equation}
\mathcal{K}_{\nu }^{\mu }=\delta _{\nu }^{\mu }-\sqrt{g^{\mu \sigma
}f_{ab}\partial _{\sigma }\phi ^{a}\partial _{\nu }\phi ^{b}}.
\end{equation}%
Here $f_{ab}$ is the non-dynamical reference metric (``fiducial metric'') and the rectangular bracket denotes the traces, namely $\left[ \mathcal{K}\right] =%
\mathcal{K}_{\mu }^{\mu }$ and $\left[ \mathcal{K}^{n}\right] =\left(\mathcal{K}^{n}\right) _{\mu }^{\mu }$. The $\phi ^{a}$'s are the St\"{u}ckelberg
scalars. Following the convention of \cite{Ghosh:2015cva}, we express $\alpha _{3}$ and $\alpha _{4}$ as
\begin{equation}
\alpha _{3}=\frac{\alpha -1}{3},~~~~~\alpha _{4}=\frac{\beta }{4}+\frac{%
1-\alpha }{12},
\end{equation}%
where $\alpha $ and $\beta $ are two arbitrary constants.

Varying the action \eqref{action} we have the field equation of this theory as
\begin{equation}
G_{\mu \nu }+m^{2}\chi _{\mu \nu }=0,  \label{Field equation}
\end{equation}%
where $G_{\mu \nu }$ is the usual Einstein tensor and the modification term $\chi _{\mu \nu }$ reads
\begin{eqnarray}
\chi _{\mu \nu } =\mathcal{K}_{\mu \nu }-\mathcal{K}g_{\mu \nu }-\alpha
\left\{ \mathcal{K}_{\mu \nu }^{2}-\mathcal{KK}_{\mu \nu }+\frac{\mathcal{U}%
_{2}}{2}g_{\mu \nu }\right\}  \notag
+3\beta ^{2}\left\{ \mathcal{K}_{\mu \nu }^{3}-\mathcal{KK}_{\mu \nu }^{2}+%
\frac{\mathcal{K}_{\mu \nu }}{2}\mathcal{U}_{2}-\frac{1}{6}g_{\mu \nu }%
\mathcal{U}_{3}\right\} .  \notag
\end{eqnarray}
Using the same choice for the nondynamical reference metric\footnote{The proof of ghost-freeness of dRGT theory \cite{HassanI,
HassanII} assumes that the reference metric is invertible, so for a degenerate reference metric like this, one has to establish ghost-freeness separately. This was done in \cite{1510.03204}. However, ghosts might still arise for some values of the black hole parameters. We shall not deal with this subtle issue in our work.} as \cite{Vegh,Cai,HendiEP1},
\begin{equation}
f_{ab}=\text{diag}(0,0,c^{2},c^{2}\sin ^{2}\theta ),  \label{reference}
\end{equation}
where $c$ is a constant with dimension of length,
we can obtain the black hole metric as
\begin{equation}
\text{d}s^{2}=-f(r)\text{d}t^{2}+\frac{\text{d}r^{2}}{f(r)}+r^{2}\left(
\text{d}\theta ^{2}+\sin ^{2}\theta \text{d}\varphi ^{2}\right),
\label{metric}
\end{equation}
where
\begin{equation}
f\left( r\right) =1-\frac{2M}{r}+\frac{r^2}{\ell^2}+\gamma r+\varepsilon.  \label{BH}
\end{equation}%
Here $M$ is an integration constant related to the black hole mass, while $\ell^2$, $\gamma$ and $\varepsilon$, are defined by
\begin{eqnarray}
\ell^2&=&\frac{1}{m^{2}\left( 1+\alpha +\beta \right)} ,  \notag \\
\gamma&=&-cm^{2}\left( 1+2\alpha +3\beta \right) ,  \notag \\
\varepsilon&=&c^{2}m^{2}\left( \alpha +3\beta \right).  \label{CondI}
\end{eqnarray}
These parameters play the roles of, respectively, the cosmological constant, linear term, and ``global monopole'' \cite{Ghosh:2015cva}. All the above three terms are contributed by the graviton mass $m$. In the massless limit $m\rightarrow0$, the black hole solution reduces to the Schwarzschild case in asymptotically flat spacetime. This solution is similar with the result of \cite{Cai}, where the cosmological constant is introduced by hand in the action, while in our case the effective cosmological constant term is \emph{emergent}, as the result of the massive graviton.

For different choices of the parameters, the black hole could have multiple horizons. Since we are considering the asymptotically AdS case\footnote{Strictly speaking, due to the presence of the linear term, the asymptotic is not strictly AdS. However since $r^2$ term dominates over $r$ in the asymptotic region, for the sake of convenience we still loosely refer to these spacetimes as asymptotically AdS, or asymptotically AdS-like.}, the black hole event horizon $r_+$ is defined as the largest root of $f(r)=0$. We can write the black hole mass as the function of $r_+$. This gives
\begin{equation}
M=\frac{r_+}{2}\left(1+\varepsilon+\frac{r_+^2}{\ell^2}+\gamma r_+ \right).
\end{equation}
The Hawking temperature is given by
\begin{align}
T=\frac{1}{4\pi r_+}\left( 1+\varepsilon+\frac{3r_+^2}{\ell^2}+2\gamma r_+ \right).
\label{temperature}
\end{align}
The Bekenstein-Hawking entropy, which can be calculated by using the first law of black hole thermodynamics $\d S=\d M/T$, yields the standard area
law $S=\pi r_+^2$.

\section{Black Hole Evaporation in dRGT massive Gravity}\label{(III)}

In the last section we have reviewed the black hole solution and thermodynamics of dRGT massive gravity. Now we are ready to investigate the black hole evaporation process. Because of the Hawking radiation, the black hole mass $M$ should be some monotonically-decreasing functions of time $t$ (we impose an absorbing boundary condition following Page). Applying the geometrical optics approximation, we assume all the emitted massless particles move along null geodesics. If we orient the angular coordinate $\varphi$ and normalize the affine parameter $\lambda$, we have the geodesic equation of the massless particles
\begin{align}
\bigg(\frac{\mathrm{d}r}{\mathrm{d}\lambda}\bigg)^2=E^2-J^2\frac{f(r)}{r^2},
\end{align}
where $E=f(r)\frac{\mathrm{d}t}{\mathrm{d}\lambda}$ is the energy and $J=r^2\frac{\mathrm{d}\theta}{\mathrm{d}\lambda}$ is the angular momentum. Consider an emitted particle from the black hole. If there is a turning point satisfying $\big(\frac{\mathrm{d}r}{\mathrm{d}\lambda}\big)^2=0$, it will turn back towards the black hole and thus cannot be detected by an observer on the AdS boundary. Defining the impact parameter as $b\equiv {J}/{E}$, the emitted particle can reach infinity if
\begin{align}
\frac{1}{b^2}> \frac{f(r)}{r^2},
\end{align}
for all $r> r_+$.

The impact factor $b_c$ can be defined by the maximal value of ${f(r)}/{r^2}$. Once we obtained $b_c$, according to the Stefan-Boltzmann law, the Hawking emission rate is
\begin{align}
\frac{\mathrm{d} M}{\mathrm{d}t}=-\mathfrak{g} Cb_c^{2} T^4,
\label{law}
\end{align}
with the constant $C=\frac{\pi^3 k^4}{15c^3 \hbar^3}$. Since we are only concerned about the qualitative features of the evaporation process, without loss of generality, we will
absorb this term into the grey-body factor $\mathfrak{g}$, which we then set to be unity: $\mathfrak{g} C=1$.
The Stefan-Boltzmann law implies that in $4$-dimensional spacetime the emission power is proportional to the $2$-dimensional cross section $b_c^2$ and the photon energy density $T^4$ in $3$-dimensional space (spatial dimension only). Since the $T^4$ term is of a higher order, the behavior of the temperature $T$, especially its asymptotic behavior, is extremely important in black hole evaporation process. Now we will investigate the black hole evaporation for various features of $T$ and $b_c$.

Solving the equation $T(r_+)=0$, we know there are two roots, which read
\begin{equation}
r_1=\frac{\ell}{3}\left(-\gamma \ell+\sqrt{\gamma^2\ell^2-3\varepsilon-3} \right)
\end{equation}
and
\begin{equation}
r_2=\frac{\ell}{3}\left(-\gamma \ell-\sqrt{\gamma^2\ell^2-3\varepsilon-3} \right).
\end{equation}
We also have $\frac{\partial M}{\partial r_+}\left(r_+=r_1\right)=\frac{\partial M}{\partial r_+}\left(r_+=r_2\right)=0$.

Similarly, solving the equation $\frac{\partial}{\partial r}\frac{f(r)}{r^2}=0$, we can also find there are two roots
\begin{equation}
r_{p1}=\frac{-\varepsilon-1+\sqrt{6M\gamma+\varepsilon^2+2\varepsilon+1}}{\gamma},
\end{equation}
and
\begin{equation}
r_{p2}=\frac{-\varepsilon-1-\sqrt{6M\gamma+\varepsilon^2+2\varepsilon+1}}{\gamma}.
\end{equation}
These correspond to the photon orbits. The maximal value $r_{p1}$ goes to $3M$ as $\varepsilon\rightarrow 0$ and $\gamma\rightarrow 0$. Furthermore, if we compare the effective potential ${f(r_{p1})}/{r_{p1}^2}$ at $r_{p1}$ with the result $1/{\ell^2}$ at infinity, we can define a critical mass
\begin{equation}
M_c=-\frac{(\varepsilon+1)^2}{8\gamma},
\end{equation}
such that for $M>M_c$ we have $\frac{1}{\ell^2}>\frac{f(r_{p1})}{r_{p1}^2}$, while for $M<M_c$ we have  $\frac{f(r_{p1})}{r_{p1}^2}>\frac{1}{\ell^2}$.

Of course the values of $r_1, r_2$, $M_c$, and $r_{p1}, r_{p2}$ depend on our choices of the coefficients, and they could be complex or correspond to results which are unphysical (for example, if there are two horizons, then like the more familiar Reissner-Nordstr\"om case in general relativity, $r$ is a timelike coordinate in between the horizon, and $r_p$ can no longer be interpreted as a photon orbit if it lies in said region). Now let us investigate the different cases in details.

In the following plots, the values of the parameters are chosen such that the interesting features can be nicely plotted. Sometimes this results in values that are ``too small'', e.g. $M < 1$ is less than a Planck mass in our units, and there is no reason to expect that black hole evaporation still obeys the usual Stefan-Boltzmann equation at the Planck scale. Our choice is therefore for convenience only. The same features are present also for reasonably larger values of the mass. In fact, such a simple evaporation model utilizing geometric optics approximation could break down much earlier before Planck mass regime is reached. In this work, we are primarily concerned with studying the difference between dRGT black hole and that of the usual Schwarzschild and Schwarzschild-AdS black hole of general relativity, assuming that the simple model holds.

\subsection{$\gamma^2\ell^2<3(\varepsilon+1)$}

\begin{figure}[h!]
\begin{center}
\includegraphics[width=0.45\textwidth]{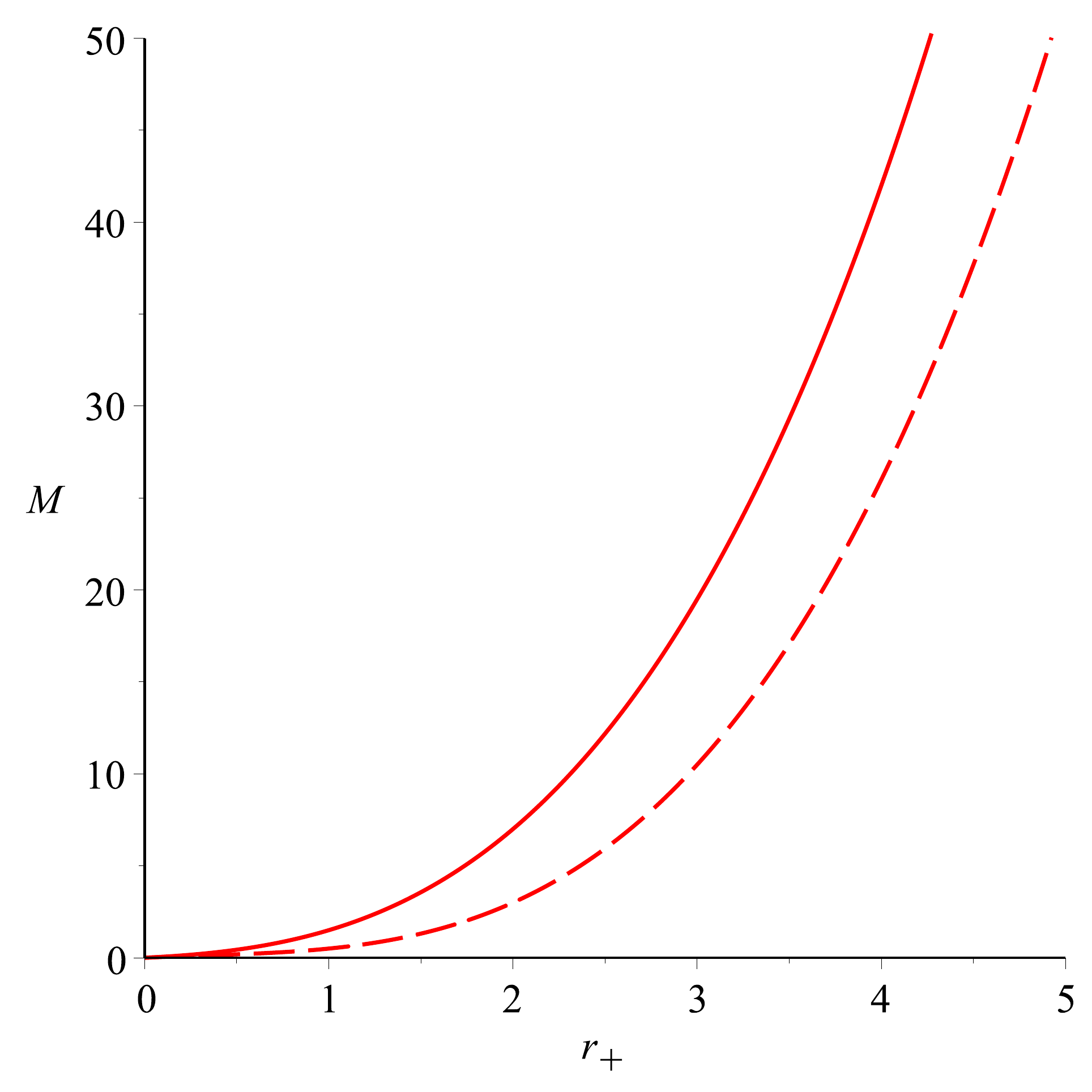}
\includegraphics[width=0.45\textwidth]{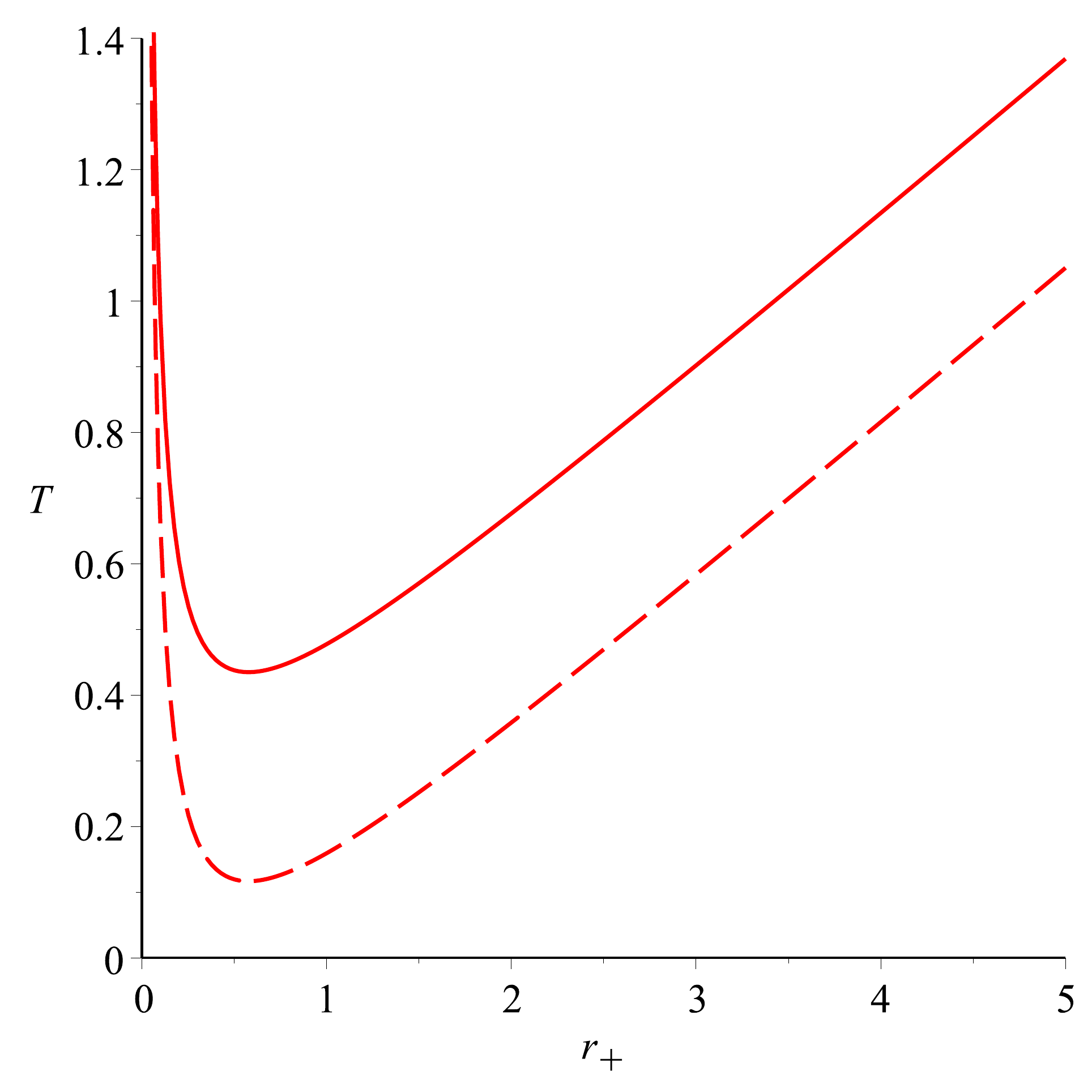}
\vspace{-1mm}
\caption{Behavior of the black hole mass $M$ and temperature $T$ as function of the horizon radius $r_+$ in case of $\gamma^2\ell^2<3(\varepsilon+1)$. We set $\varepsilon=0$, $\ell=1$. The solid and dashed curves correspond to $\gamma=1$ and $\gamma=-1$ respectively.}
\label{fig1}
\end{center}
\end{figure}

Let us first consider the case in which both $r_1$ and $r_2$ are unreal\footnote{By unreal we mean it is an element of $\Bbb{C}\setminus\Bbb{R}$.} roots. This demands $\gamma^2\ell^2<3(\varepsilon+1)$. In this case the temperature never vanishes. The behavior of the temperature $T$ resembles the case of AdS-Schwarzschild with Hawking-Page phase transition, and the black hole mass is a monotonic function of $r_+$. The coefficient $\gamma$ can be positive or negative, and the sign of $\gamma$ does not affect the qualitative features of $M$ and $T$ in this case. In FIG.\ref{fig1} we present some examples.

\begin{figure}[h!]
\begin{center}
\includegraphics[width=0.45\textwidth]{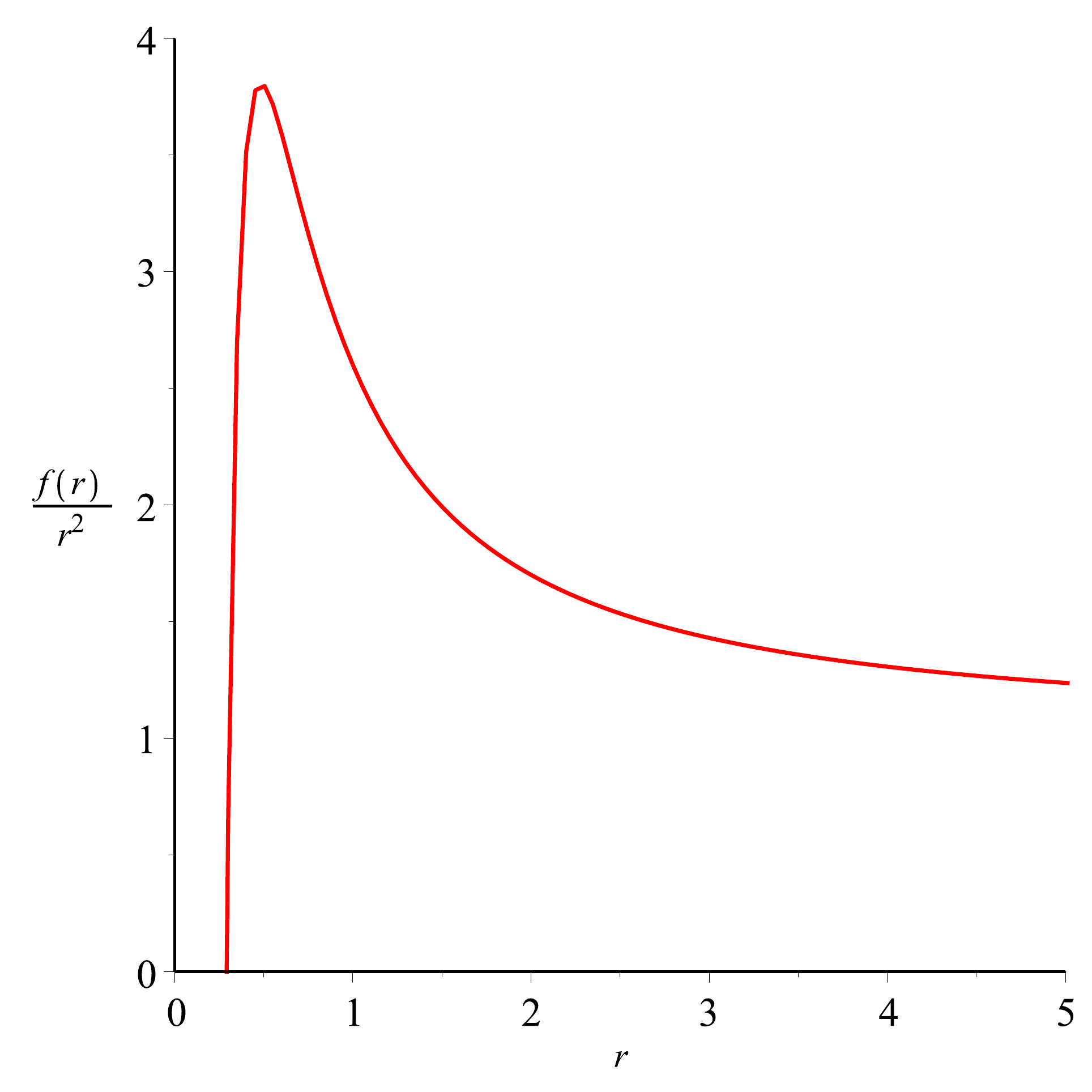}
\includegraphics[width=0.45\textwidth]{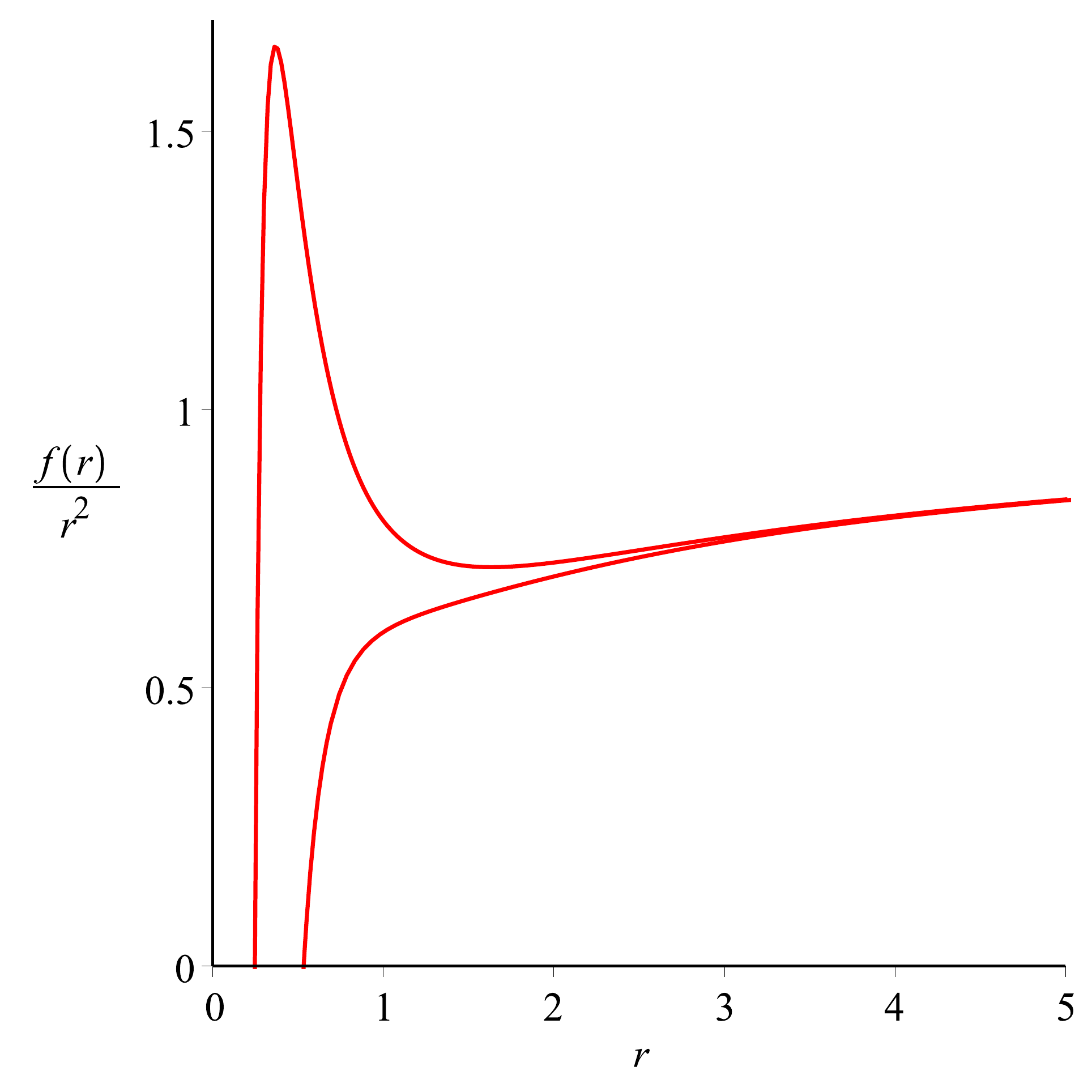}
\vspace{-1mm}
\caption{Behavior of the effective potential ${f(r)}/{r^2}$ as function of $r$ in the case $\gamma^2\ell^2<3(\varepsilon+1)$. In the left figure we set $\varepsilon=0$, $\ell=1$, $\gamma=1$ and $M={1}/{5}$. In the right figure $\varepsilon=0$, $\ell=1$, $\gamma=-1$, while the upper and lower curves correspond to $M={1}/{10}<M_c$ and $M={1}/{5}>M_c$, respectively.}
\label{fig2}
\end{center}
\end{figure}

However, the sign of $\gamma$ affects the features of the effective potential $\frac{f(r)}{r^2}$. For the case of $\gamma>0$, the effective potential admits its maximal value at $r=r_{p1}=\frac{-\varepsilon-1+\sqrt{6M\gamma+\varepsilon^2+2\varepsilon+1}}{\gamma}$, and the impact factor $b_c=\frac{r_{p1}}{\sqrt{f(r_{p1})}}$. There is no critical mass $M_c$. For the case of $\gamma<0$, on the other hand, $r_{p1}$ and $r_{p2}$ correspond to the maximal and minimal value of effective potential respectively\footnote{At the minimum of the potential, the photon orbit is stable, which indicates that the spacetime might be unstable (because backreaction of massless particles accumulating on said orbit would modify the black hole metric); see also \cite{1406.5510}. However in this work we do not consider all the various ways that the black holes might be unstable, and only focus on the Hawking process.}. For $M>M_c$, the impact parameter is $b_c=\ell$. For $M<M_c$, the impact parameter becomes $b_c=\frac{r_{p1}}{\sqrt{f(r_{p1})}}$. We present some examples of $\frac{f(r)}{r^2}$ in FIG.\ref{fig2}.

Now we can investigate the black hole evaporation using Stefan-Boltzmann law. By scaling analysis we know $M\sim l$, $T\sim l^{-1}$ and $b_c\sim l$, where $l$ denotes some length. However, note that the linear coefficient $\gamma$ also scales as $\gamma\sim l^{-1}$. Defining the dimensionless variables\footnote{Of course in our units everything is dimensionless. The point is the quantities $x$ and $y$ are dimensionless in any unit one may choose.} $x\equiv {r_+}/{\ell}$ and $y\equiv \gamma \ell$, we can express $M$, $T$ and $r_{p1}$ as
\begin{equation}
M=\mathcal{M}(x,y,\varepsilon)\ell=\frac{x}{2}\left(1+\varepsilon+x^2+xy \right)\ell,
\label{M}
\end{equation}
\begin{equation}
T=\mathcal{T}(x,y,\varepsilon)\ell^{-1}=\frac{1}{4\pi x}\left(1+\varepsilon +3x^2+2xy \right)\ell^{-1},
\label{T}
\end{equation}
\begin{equation}
r_{p1}=\mathcal{R}(x,y,\varepsilon)\ell=\left(-\varepsilon-1+\sqrt{3xy(1+\varepsilon+x^2+xy)+(\varepsilon+1)^2}\right)\frac{1}{y}\ell,
\end{equation}
where $\mathcal{M}(x,y,\varepsilon)$, $\mathcal{T}(x,y,\varepsilon)$, $\mathcal{R}(x,y,\varepsilon)$ are all dimensionless functions. Inserting the above $M$ and $r_{p1}$ into $b_c=\frac{r_{p1}}{\sqrt{f(r_{p1})}}$, we find that the impact parameter $b_c$ can also be written as
\begin{equation}
b_c=\mathcal{B}(x,y,\varepsilon)\ell=\frac{\mathcal{R}(x,y,\varepsilon)}{\sqrt{1+\varepsilon-\frac{2\mathcal{M}(x,y,\varepsilon)}{\mathcal{R}(x,y,\varepsilon)}+\mathcal{R}^2(x,y,\varepsilon)+\mathcal{R}(x,y,\varepsilon)y}}\ell.
\label{bc}
\end{equation}

In the case of $\gamma>0$, the black hole impact factor is $b_c=\frac{r_{p1}}{\sqrt{f(r_{p1})}}$. Inserting equations \eqref{M}, \eqref{T}, and \eqref{bc} into the Stefan-Boltzmann law, we have
\begin{equation}
\d t=-\frac{\partial \mathcal{M}(x,y,\varepsilon)}{\partial x}\frac{\ell^3}{\mathcal{B}^2(x,y,\varepsilon)\mathcal{T}^4(x,y,\varepsilon)}\d x.
\end{equation}
If we set $y$ and $\varepsilon$ to be constant, we can obtain the black hole lifetime by integrating the above formula from $\infty$ to $0$. This integration turns out to be finite, and the black hole lifetime is of the order $\ell^3$.

For the case of $\gamma<0$, there is a critical mass $M_c$. For $M>M_c$ the impact factor is $b_c=\ell$, while for $M<M_c$ it is $b_c=\frac{r_{p1}}{\sqrt{f(r_{p1})}}$. Solving the equation $M=M_c$ we can obtain the corresponding $x_c$ satisfying
\begin{equation}
\mathcal{M}(x_c,y,\varepsilon)=-\frac{(1+\varepsilon)^2}{8y},
\end{equation}
which depends on the values of $y$ and $\varepsilon$. Once we set $y$ and $\varepsilon$ to be constant, $x_c$ is also fixed. Applying the Stefan-Boltzmann law, the lifetime of an arbitrarily large black hole reads
\begin{equation}
t=-\int^{\infty}_{x_c}\frac{\partial \mathcal{M}(x,y,\varepsilon)}{\partial x}\frac{\ell^3}{\mathcal{T}^4(x,y,\varepsilon)}\d x-\int^{x_c}_{0}\frac{\partial \mathcal{M}(x,y,\varepsilon)}{\partial x}\frac{\ell^3}{\mathcal{B}^2(x,y,\varepsilon)\mathcal{T}^4(x,y,\varepsilon)}\d x.
\end{equation}
Thus we see that the black hole lifetime is still of the order $\ell^3$.

\begin{figure}[h!]
\begin{center}
\includegraphics[width=0.48\textwidth]{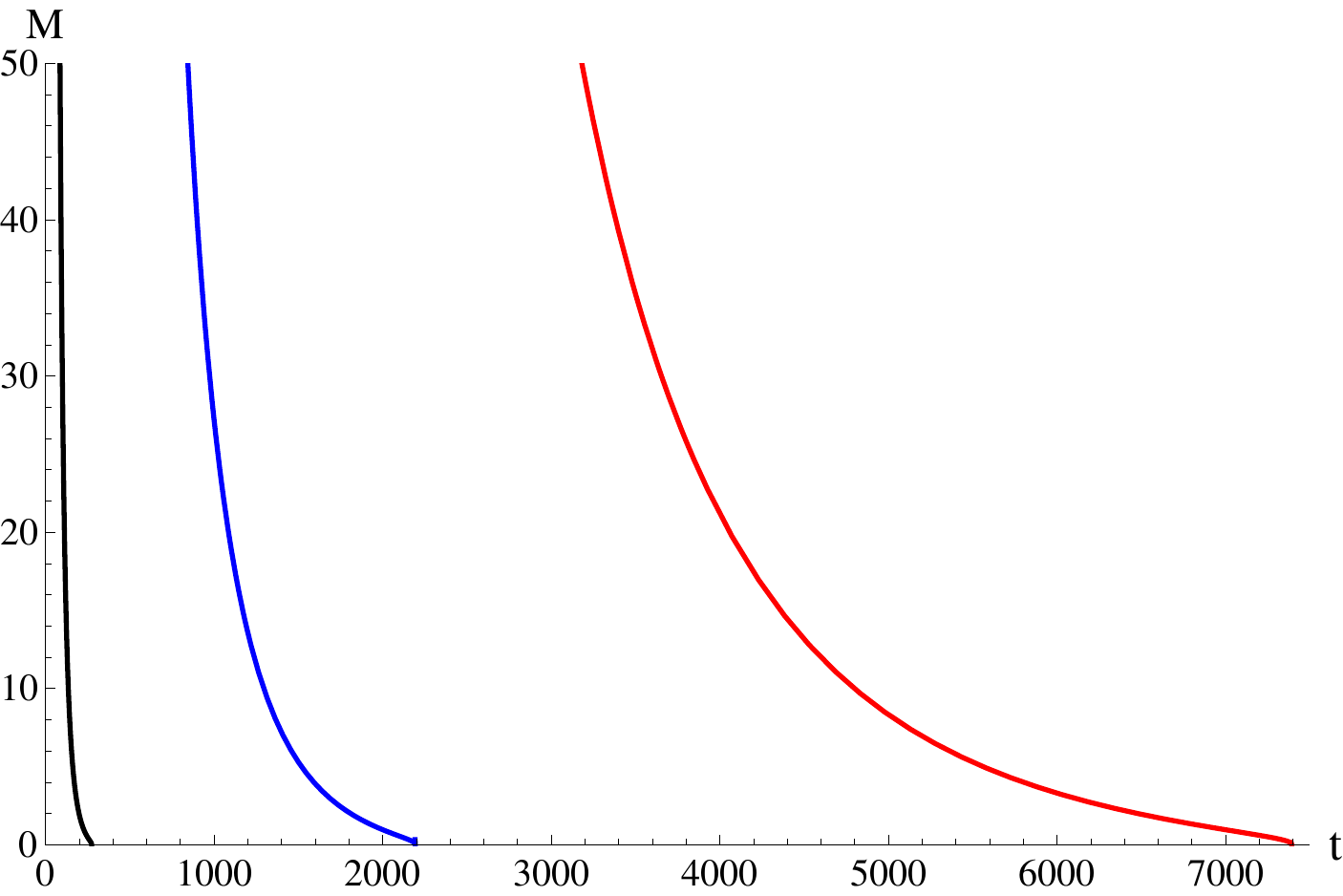}
\includegraphics[width=0.48\textwidth]{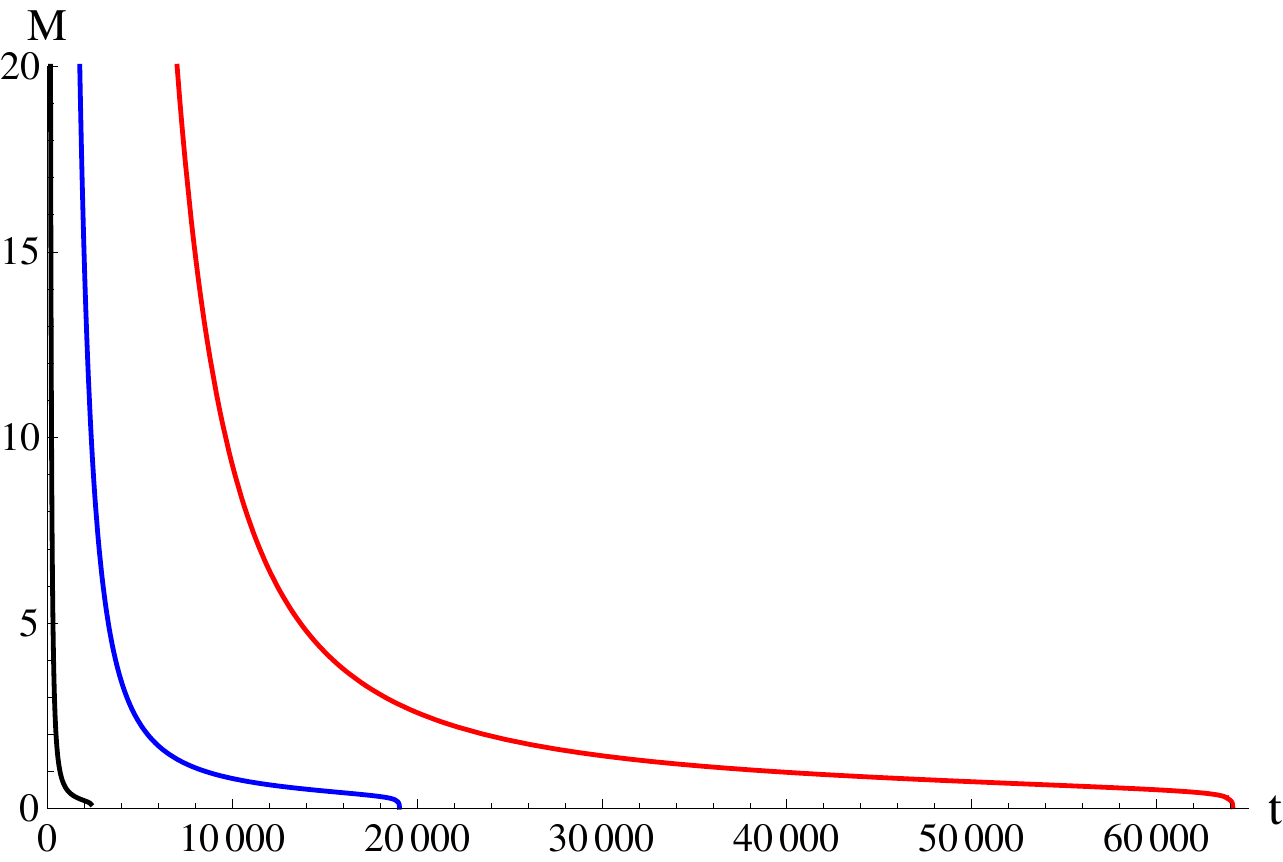}
\vspace{-1mm}
\caption{The evolution of the black hole for $\gamma^2\ell^2<3(\varepsilon+1)$. For the left figure we have $\varepsilon=0$, $y=\gamma \ell=1$. For the right figure we have $\varepsilon=0$, $y=\gamma \ell=-1$. In each figure from left to right the curves correspond to $\ell=1$, $\ell=2$ and $\ell=3$ respectively.}
\label{fig3}
\end{center}
\end{figure}

In FIG.\ref{fig3} we present the evolution of the black hole for $\gamma^2\ell^2<3(\varepsilon+1)$. For the left figure we choose $\varepsilon=0$, $y=\gamma \ell=1$, and for the right figure we choose $\varepsilon=0$, $y=\gamma \ell=-1$. In each figure from left to right the curves correspond to $\ell=1$, $\ell=2$ and $\ell=3$ respectively. We find that the black hole lifetime is always finite, and it is of the order $\ell^3$.

\subsection{$\gamma^2\ell^2>3(\varepsilon+1)$, $\gamma>0$, $\varepsilon+1>0$}

\begin{figure}[h!]
\begin{center}
\includegraphics[width=0.48\textwidth]{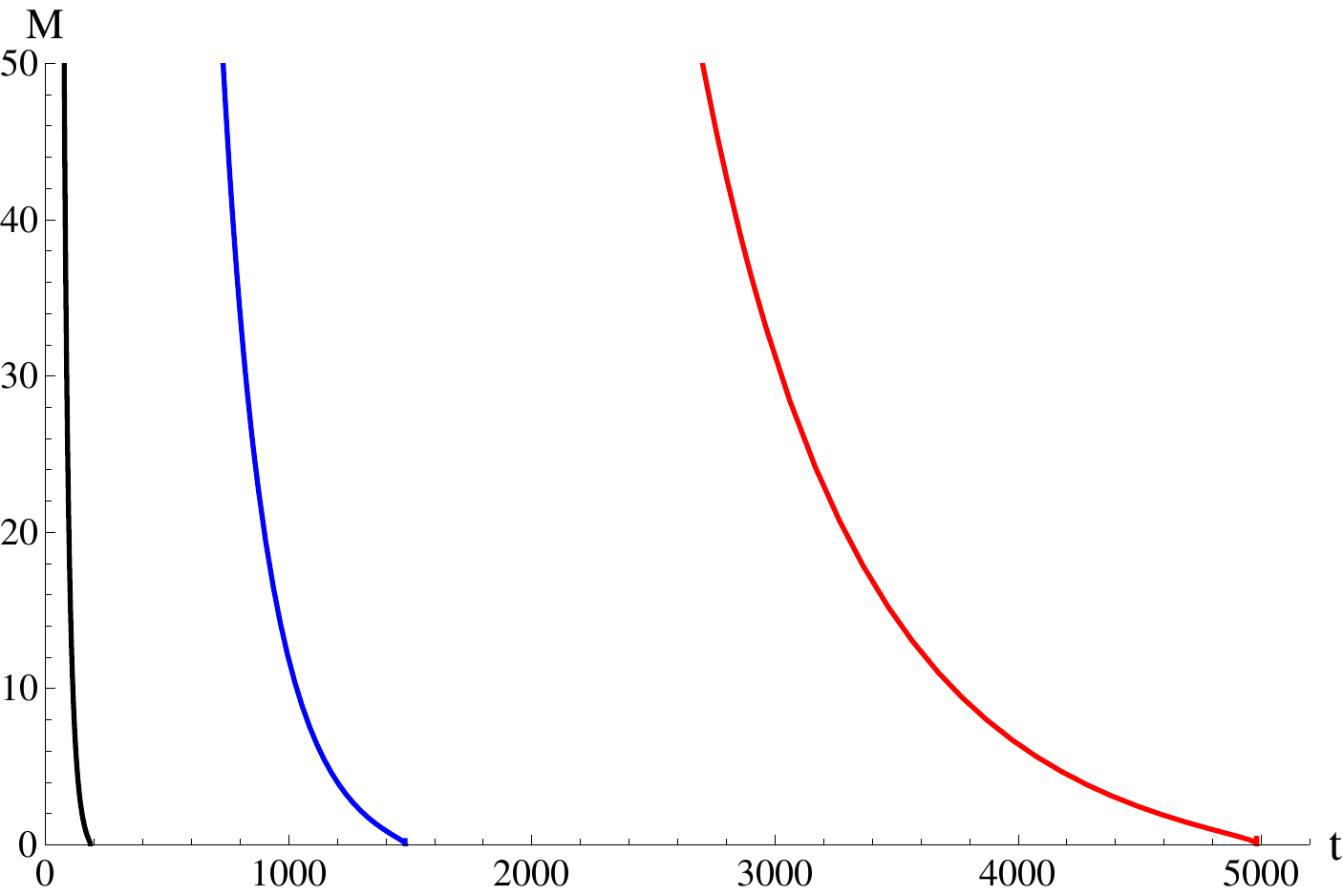}
\vspace{-1mm}
\caption{The evolution of the black holes for $\gamma^2\ell^2>3(\varepsilon+1)$, $\gamma>0$, $\varepsilon+1>0$. We set $\gamma\ell=2$, $\varepsilon=0$. From left to right the curves correspond to $\ell=1$, $\ell=2$ and $\ell=3$ respectively.}
\label{fig4}
\end{center}
\end{figure}

For the second case, we shall consider the situation that the roots $r_1$ and $r_2$ are both real but negative. This demands $\gamma^2\ell^2>3(\varepsilon+1)$, $\gamma>0$, $\varepsilon+1>0$. For the physical domain $r_+>0$, the black hole mass is a monotonic function of $r_+$. The Hawking temperature $T$ again resembles the case with Hawking-Page phase transition, which is similar to the cases in FIG.\ref{fig1}. The effective potential admits a maximal value at $r_{p1}$. We always have $b_c=\frac{r_{p1}}{\sqrt{f(r_{p1})}}$. Again from Stefan-Boltzmann law, we have
\begin{equation}
\d t=-\frac{\partial \mathcal{M}(x,y,\varepsilon)}{\partial x}\frac{\ell^3}{\mathcal{B}^2(x,y,\varepsilon)\mathcal{T}^4(x,y,\varepsilon)}\d x.
\end{equation}
Setting $y$ and $\varepsilon$ to be constant and integrating the above formula from $\infty$ to $0$, we can also check that this integration is finite and the black hole lifetime is again of the order $\ell^3$. In FIG.\ref{fig4} we present some examples of the black hole evolution. We set $\gamma\ell=2$, $\varepsilon=0$, and from left to right the curves correspond to $\ell=1$, $\ell=2$ and $\ell=3$ respectively.

\subsection{$\gamma^2\ell^2>3(\varepsilon+1)$, $\varepsilon+1<0$}

\begin{figure}[h!]
\begin{center}
\includegraphics[width=0.45\textwidth]{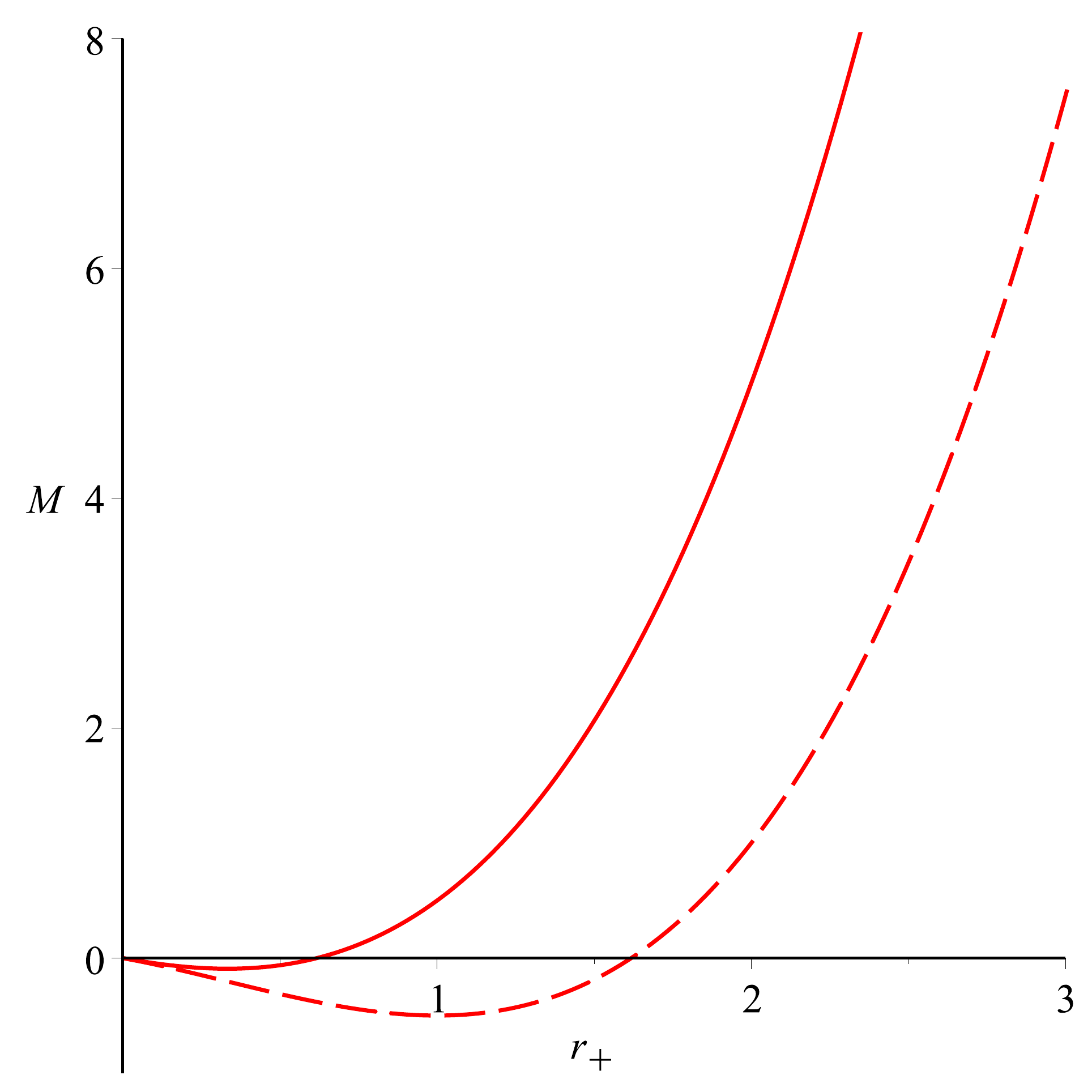}
\includegraphics[width=0.45\textwidth]{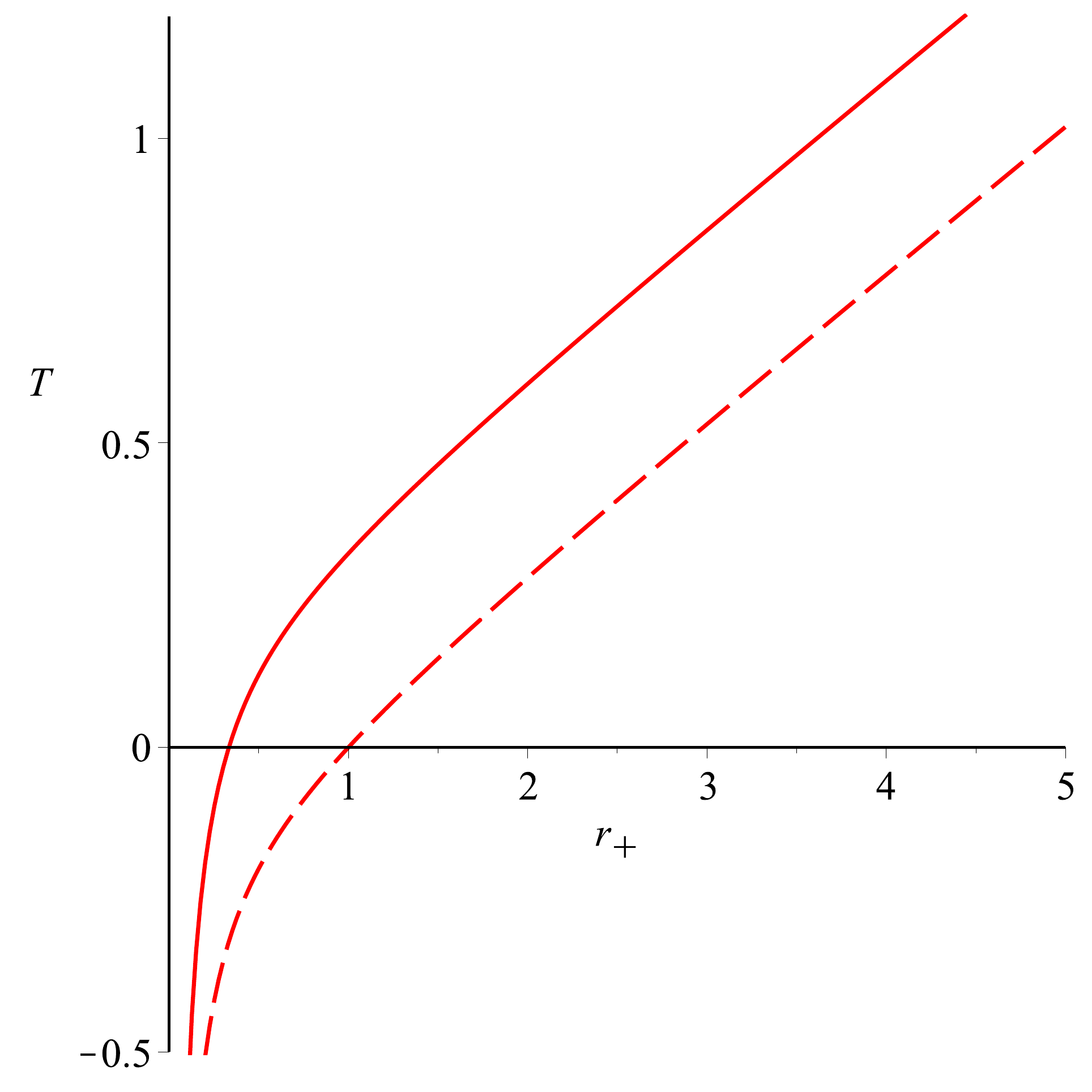}
\vspace{-1mm}
\caption{Behavior of the black hole mass $M$ and temperature $T$ as functions of the horizon radius $r_+$. We set $\varepsilon=-2$, $\ell=1$. The solid and dashed curves correspond to $\gamma=1$ and $\gamma=-1$ respectively. Negative temperature is unphysical, but negative mass is not necessarily unphysical.}
\label{fig5}
\end{center}
\end{figure}

For the third case, we consider the case that both the roots $r_1$ and $r_2$ are real, but only the larger root $r_1$ is positive. This demands $\gamma^2\ell^2>3(\varepsilon+1)$, $\varepsilon+1<0$. This situation is qualitatively different from the two cases we discussed above. The temperature $T$ and $\frac{\partial M}{\partial r_+}$ vanish at $r_+=r_1$. Since the mass $M\rightarrow 0$ as $r_+\rightarrow 0$, and $M\rightarrow \infty$ as $r_+\rightarrow\infty$, the root $r_1$ corresponds to a minimal value of $M$ which is negative. This is due to the global monopole term $\varepsilon+1<0$ in our case, so the metric function is similar to the hyperbolic Schwarschild-AdS black hole. We cannot yet conclude that $M<0$ is unphysical since the ground state may not be $M=0$ (for the hyperbolic Schwarzschild-AdS case, see \cite{9705004, 9906040} for further discussion).
In FIG.\ref{fig5} we present some examples of the behavior of the black hole mass $M$ and temperature $T$ as function of the horizon radius $r_+$. The solid and dashed curves correspond to $\gamma=1$ and $\gamma=-1$ respectively. For completeness, as well as to aid understanding, we present the whole curves, but it should be emphasized that the region in which $T<0$ and $\frac{\partial M}{\partial r_+}<0$ need to be excluded as they are not physical.

\begin{figure}[h!]
\begin{center}
\includegraphics[width=0.45\textwidth]{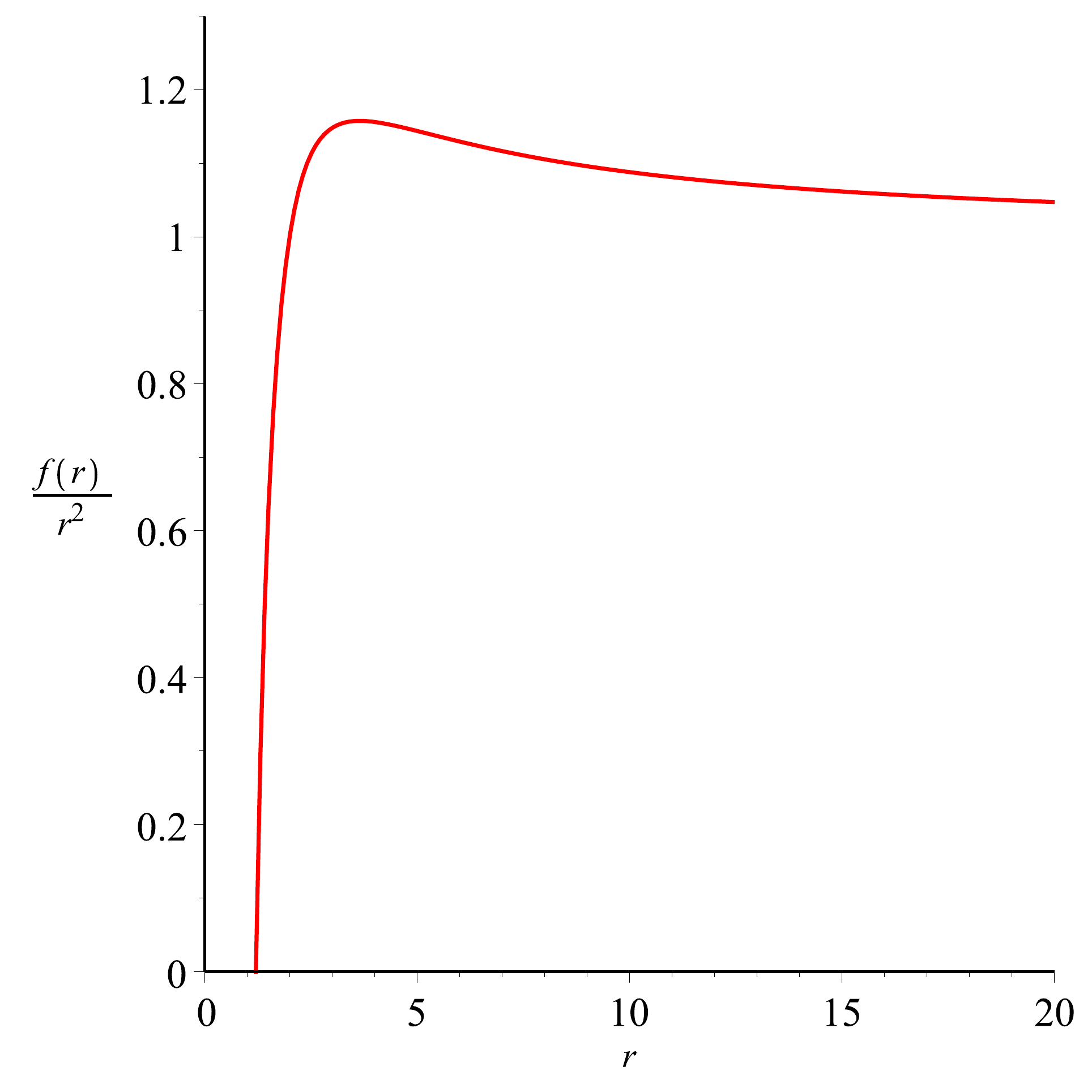}
\includegraphics[width=0.45\textwidth]{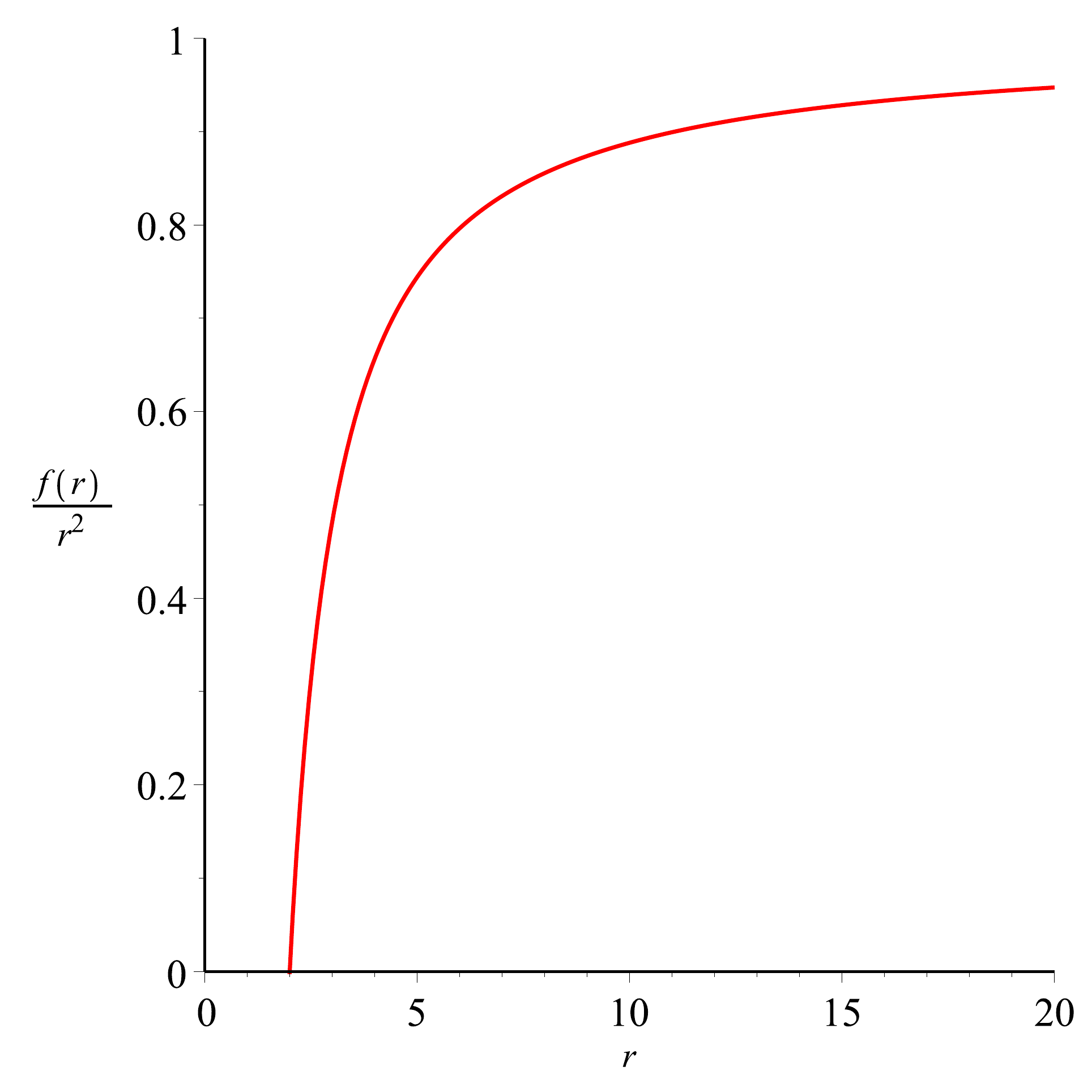}
\vspace{-1mm}
\caption{Behavior of the effective potential $\frac{f(r)}{r^2}$ as function of $r$. In the left figure we set $\varepsilon=-2$, $\ell=1$, $\gamma=1$ and $M=1$. In the right figure $\varepsilon=-2$, $\ell=1$, $\gamma=-1$, $M=1$.}
\label{fig6}
\end{center}
\end{figure}

For different sign of $\gamma$ the qualitative features of the effective potential are also different. See FIG.\ref{fig6} for examples. For $\gamma>0$, we have $b_c=\frac{r_{p1}}{\sqrt{f(r_{p1})}}$, while for $\gamma<0$, we have $b_c=\ell$. For $M<0$ the term $\frac{f(r)}{r^2}$ admits a minimal value inside the horizon, which is not of physical relevance. Again, applying the Stefan-Boltzmann law, we have
\begin{equation}
\d t=-\frac{\partial \mathcal{M}(x,y,\varepsilon)}{\partial x}\frac{\ell^3}{\mathcal{B}^2(x,y,\varepsilon)\mathcal{T}^4(x,y,\varepsilon)}\d x
\end{equation}
for $\gamma>0$, and
\begin{equation}
\d t=-\frac{\partial \mathcal{M}(x,y,\varepsilon)}{\partial x}\frac{\ell^3}{\mathcal{T}^4(x,y,\varepsilon)}\d x
\end{equation}
for $\gamma<0$. However, in both cases we always have $\mathcal{T}(x,y,\varepsilon)=0$ at $x=\frac{r_1}{\ell}=\frac{1}{3}\left(-y+\sqrt{y^2-3\varepsilon-3} \right)$, so the integration from any initial $x$ down to $x=\frac{1}{3}\left(-y+\sqrt{y^2-3\varepsilon-3} \right)$ is always divergent. The black hole can lose away a huge amount of mass from arbitrarily large initial mass to a finite mass within a finite time. However the evaporation becomes increasingly difficult when it gets near to the $T=0$ state, as expected. The black hole effectively becomes a remnant \cite{chen}. Note that this phenomenon also obeys the third law of black hole thermodynamics (the final asymptotic remnant state is an extremal black hole).

\begin{figure}[h!]
\begin{center}
\includegraphics[width=0.45\textwidth]{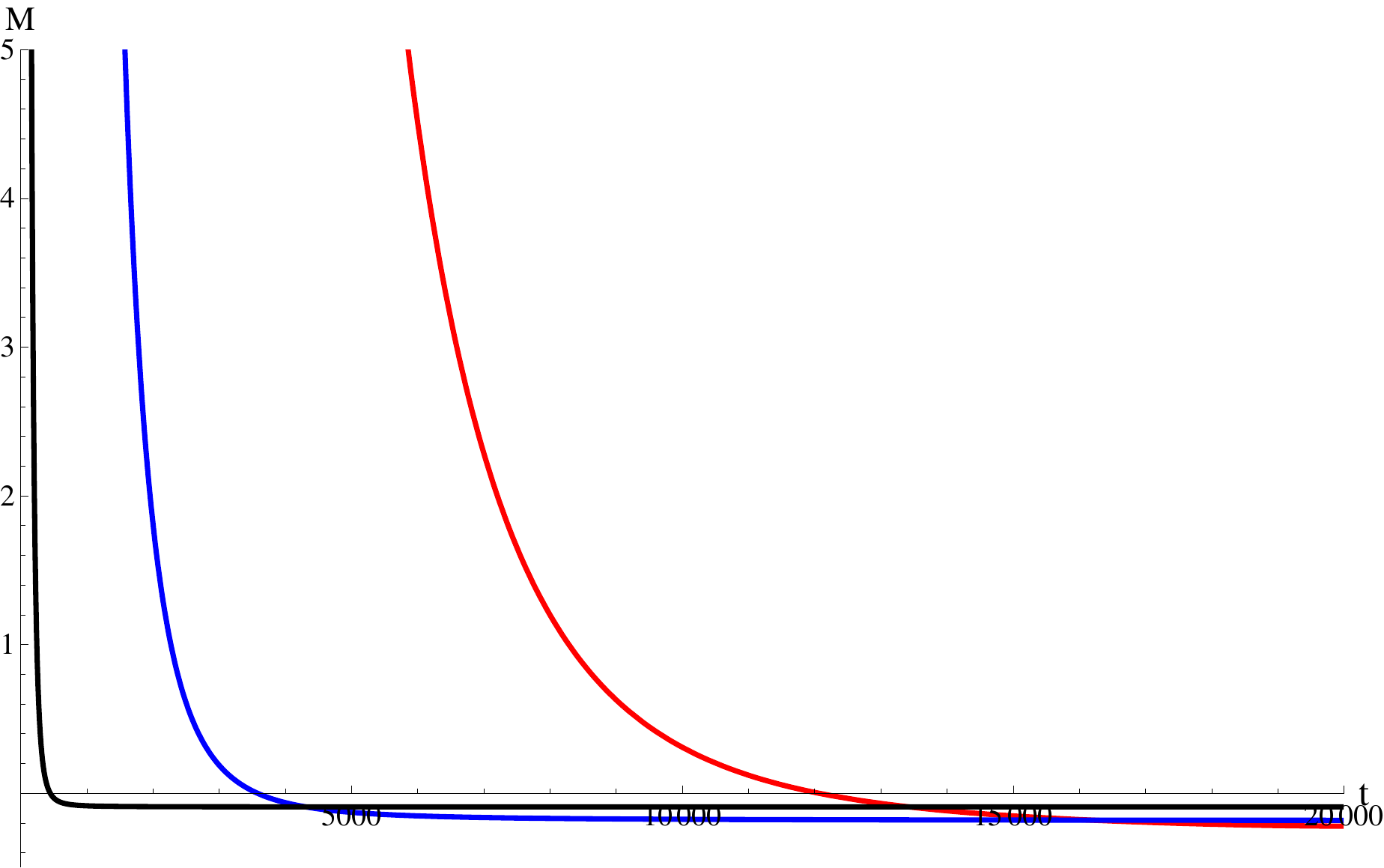}
\includegraphics[width=0.45\textwidth]{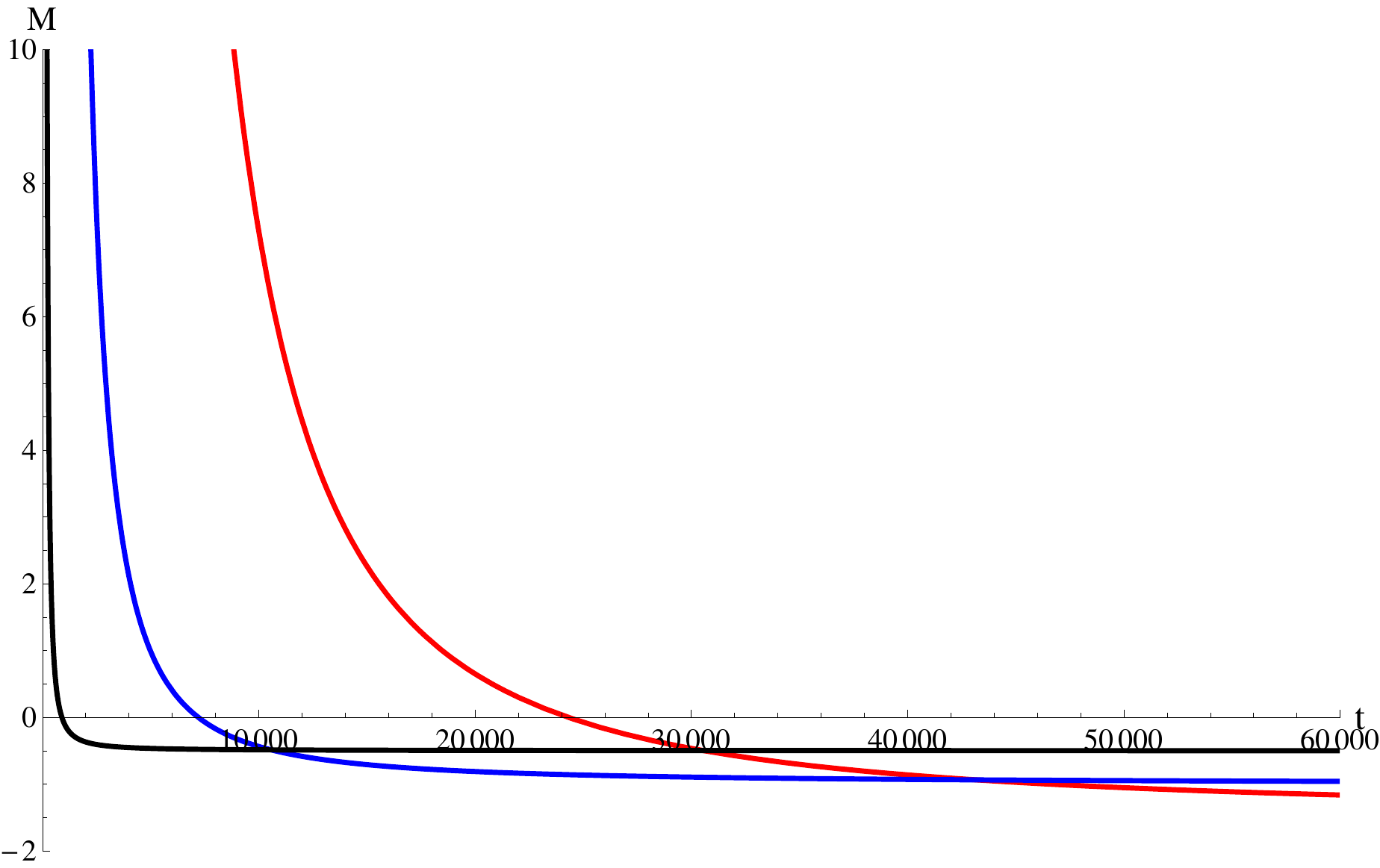}
\vspace{-1mm}
\caption{The evolution of the black hole for the case of $\gamma^2\ell^2>3(\varepsilon+1)$, $\varepsilon+1<0$. For the left figure we have $\varepsilon=-2$, $y=\gamma \ell=1$. For the right figure we have $\varepsilon=-2$, $y=\gamma \ell=-1$. In each figure from left to right the curves correspond to $\ell=1$, $\ell=2$ and $\ell=3$ respectively.}
\label{fig7}
\end{center}
\end{figure}

In FIG.\ref{fig7} we present the examples of black hole evaporation. We set $\varepsilon=-2$. For both cases of $y=\pm 1$ and various choices of $\ell$, the black holes have infinite lifetime.

\subsection{$\gamma^2\ell^2>3(\varepsilon+1)$, $\varepsilon+1>0$, $\gamma<0$}

\begin{figure}[h!]
\begin{center}
\includegraphics[width=0.45\textwidth]{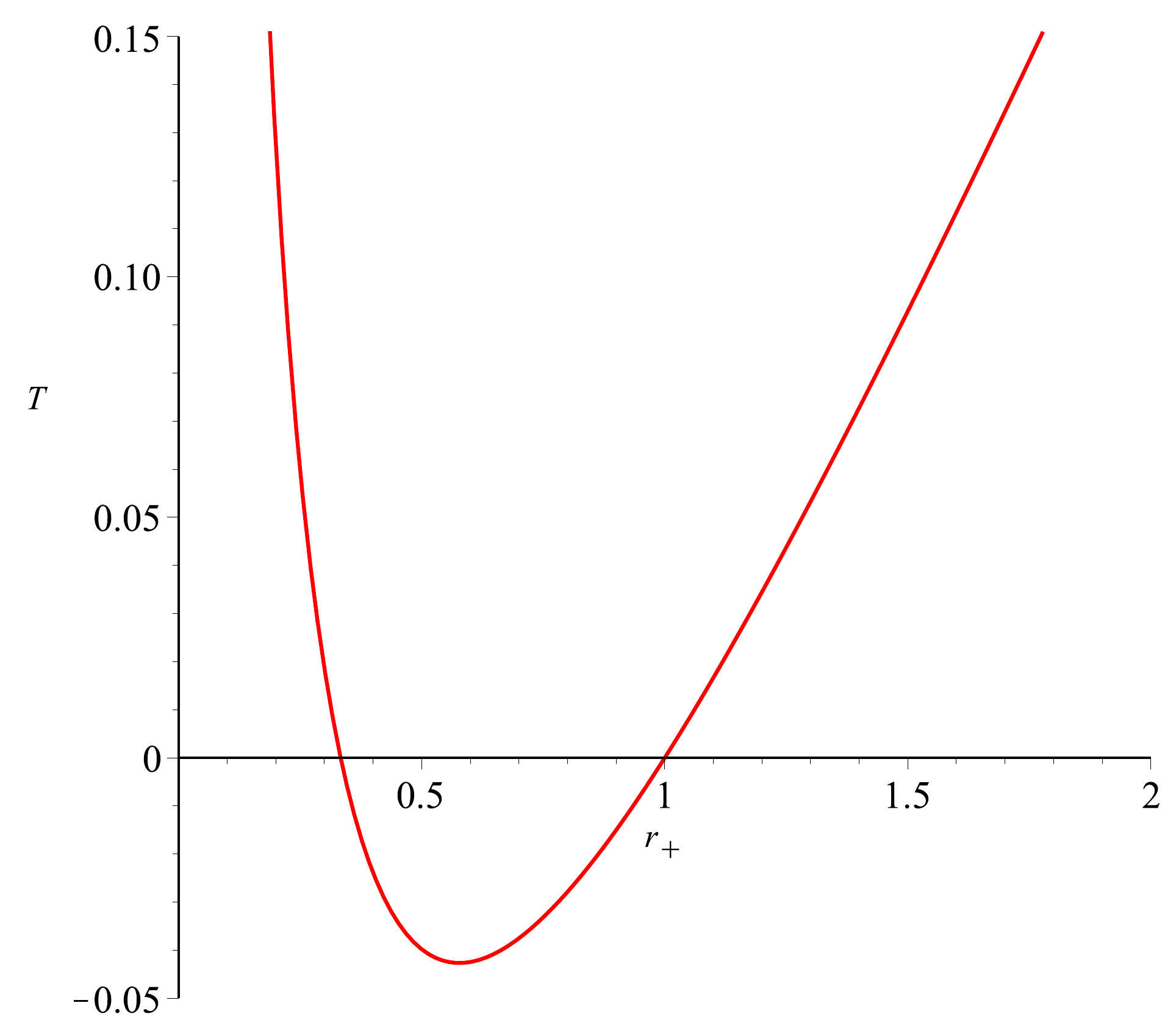}
\vspace{-1mm}
\caption{Behavior of the temperature $T$ as function of the horizon radius $r_+$ in $\gamma^2\ell^2>3(\varepsilon+1)$, $\varepsilon+1>0$, $\gamma<0$. We set $\varepsilon=0$, $\gamma=-2$, $\ell=1$.}
\label{fig8}
\end{center}
\end{figure}

For the last case, let us consider the case such that both $r_1$ and $r_2$ are real and positive.
This turned out to be the most complicated case since we have discuss 3 different mass ranges. In general
the temperature behaves like the example in FIG.\ref{fig8}. Here we present the whole curve of the temperature $T$ as the function of $r_+$ for clarity, but negative $T$ is not physical. The temperature $T\rightarrow \infty$ as $r_+\rightarrow 0$ and $r_+\rightarrow \infty$, and $T$ vanishes at $r_1$ and $r_2$. The exact values of $\varepsilon$ and $\gamma$ do not change the qualitative features of $T$, but they affect the black hole mass $M$. In the following we will consider the features of $M$ in three different ranges.

\subsubsection{$0<M(r_1)<M_c<M(r_2)$}

\begin{figure}[h!]
\begin{center}
\includegraphics[width=0.45\textwidth]{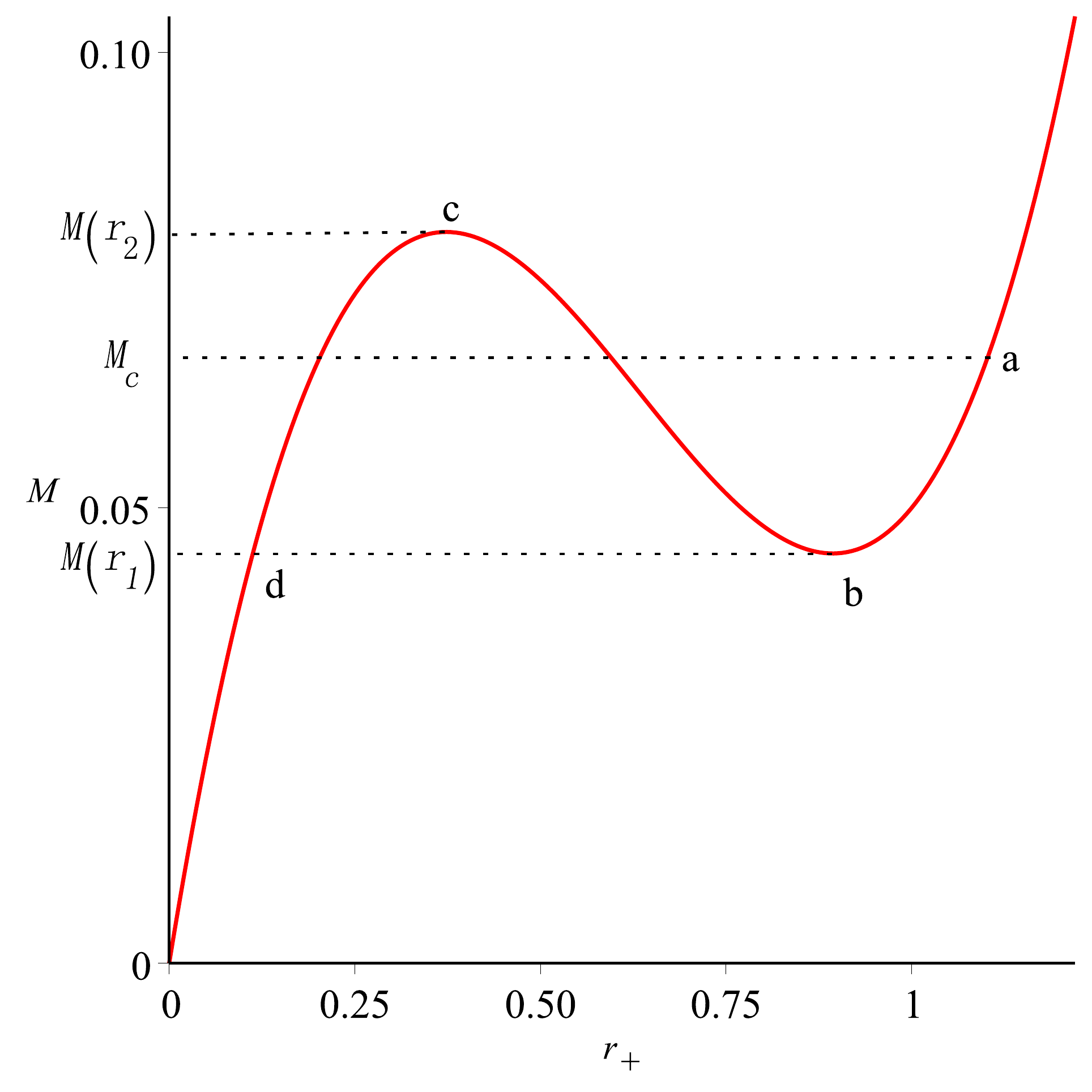}
\vspace{-1mm}
\caption{Behavior of the black hole mass $M$ as function of the horizon radius $r_+$ in $-2\sqrt{\varepsilon+1}<\gamma\ell<-{3\sqrt{6(\varepsilon+1)}}/{4}$, $\varepsilon+1>0$, $\gamma<0$. We set $\varepsilon=0$, $\gamma=-1.9$, and $\ell=1$. We have $0<M(r_1)<M_c<M(r_2)$.}
\label{fig9}
\end{center}
\end{figure}

\begin{figure}[h!]
\begin{center}
\includegraphics[width=0.45\textwidth]{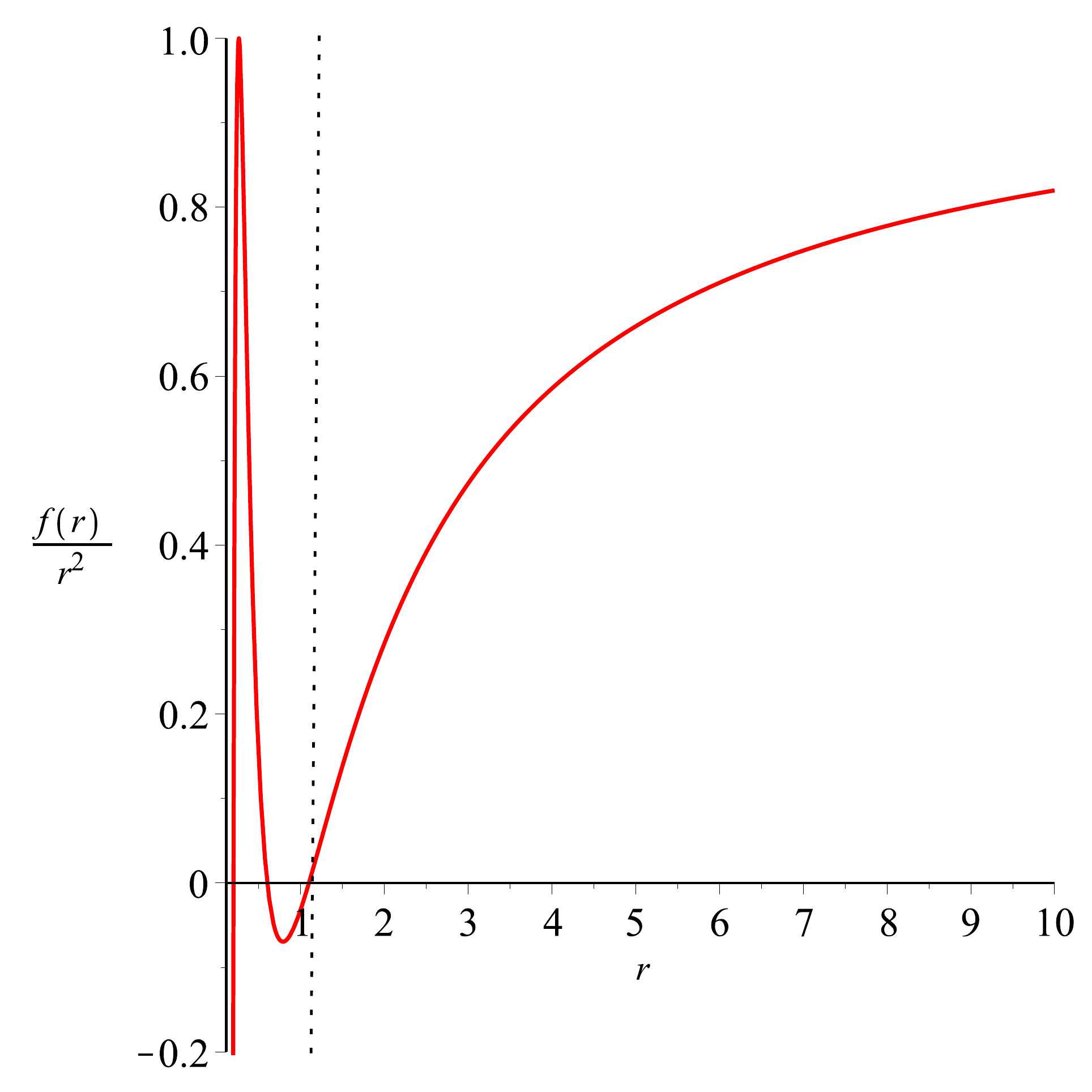}
\vspace{-1mm}
\caption{Behavior of $\frac{f(r)}{r^2}$ as function of $r$ for which $-2\sqrt{\varepsilon+1}<\gamma\ell<-{3\sqrt{6(\varepsilon+1)}}/{4}$, $\varepsilon+1>0$, $\gamma<0$. We set $\varepsilon=0$, $\gamma=-1.9$, $\ell=1$, and $M=M_c\approx 0.0658$. The dashed line corresponds to the black hole horizon. One can check that the maximal value of ${f(r)}/{r^2}$ is inside the black hole horizon.}
\label{fig10}
\end{center}
\end{figure}

Substituting $r_+=r_2$ into $M(r_+)$ it is easy to verify that $M(r_2)$ is always positive and larger than the critical mass $M_c$. Firstly we consider the case for which $M_c$ is larger that $M(r_1)$, and $M(r_1)$ is positive. A positive $M(r_1)$ requires
\begin{equation}
-2\sqrt{\varepsilon+1}<\gamma\ell<-\sqrt{3(\varepsilon+1)},
\end{equation}
and in order to have $M_c>M(r_1)$ we need
\begin{equation}
\gamma\ell<-\frac{3\sqrt{6(\varepsilon+1)}}{4}.
\end{equation}
Putting the above two conditions together we have
\begin{equation}
-2\sqrt{\varepsilon+1}<\gamma\ell<-\frac{3\sqrt{6(\varepsilon+1)}}{4}.
\end{equation}
In FIG.\ref{fig9} we present the behavior of $M$ as the function of $r_+$. We set $\varepsilon=0$, $\gamma=-1.9$, $\ell=1$. In this situation $0<M(r_1)<M_c<M(r_2)$. There are two points worth noting here. Firstly, for the case of $M(r_1)<M<M(r_2)$, since the black hole event horizon is defined as the largest root, the region \textsf{b-c-d} should be excluded, thus the solution is branched. One branch is from $M\rightarrow \infty$ to point \textsf{b}, and the other is from point \textsf{d} to $M=0$. Secondly, for $M(r_1)<M<M_c$ the term $\frac{f(r)}{r^2}$ admits a maximal value that is larger than $\frac{1}{\ell^2}$, but this value is \emph{inside} the horizon radius $r_+$, so it is not a part of the effective potential and the impact factor is still $b_c=\ell$. See FIG.\ref{fig10} for example. For $0<M<M(r_1)$ the maximal value is outside of the horizon, so we have $b_c=\frac{r_{p1}}{\sqrt{f(r_{p1})}}$.

In conclusion, the black hole solution is branched. One is from $M\rightarrow \infty$ to point \textsf{b}, and the impact factor $b_c=\ell$. One is from point d to $M=0$, and the impact factor is $b_c=\frac{r_{p1}}{\sqrt{f(r_{p1})}}$. We can study the black hole evaporation for each branch. For the first branch the initial mass can be taken to be arbitrarily large, while for the second branch, the initial mass should be slightly smaller than the $M(r_1)$.

For the first branch we have
\begin{equation}
\d t=-\frac{\partial \mathcal{M}(x,y,\varepsilon)}{\partial x}\frac{\ell^3}{\mathcal{T}^4(x,y,\varepsilon)}\d x.
\end{equation}
The finial state corresponds to $T=0$ so the lifetime of the black hole is always infinite. For the second branch we have
\begin{equation}
\d t=-\frac{\partial \mathcal{M}(x,y,\varepsilon)}{\partial x}\frac{\ell^3}{\mathcal{B}^2(x,y,\varepsilon)\mathcal{T}^4(x,y,\varepsilon)}\d x.
\end{equation}
The initial $x$ for the point d only depends on the value of $\varepsilon$ and $y$, thus for fixed $\varepsilon$ and $y$ the initial $x$ is a constant. The integration from the initial $x$ at point d to $0$ is finite, so the lifetime of the black hole is in the order of $\ell^3$.

\begin{figure}[h!]
\begin{center}
\includegraphics[width=0.45\textwidth]{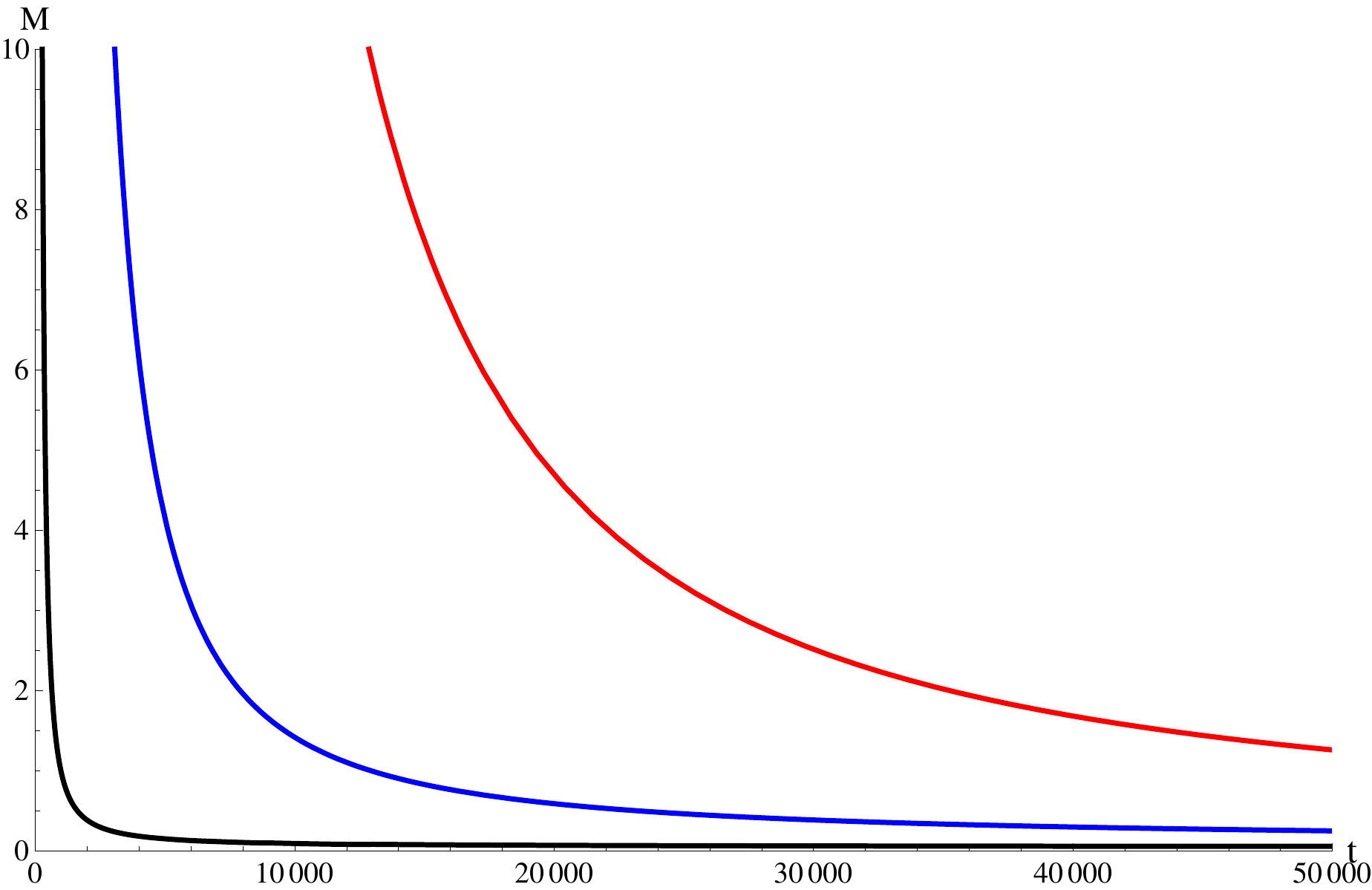}
\includegraphics[width=0.45\textwidth]{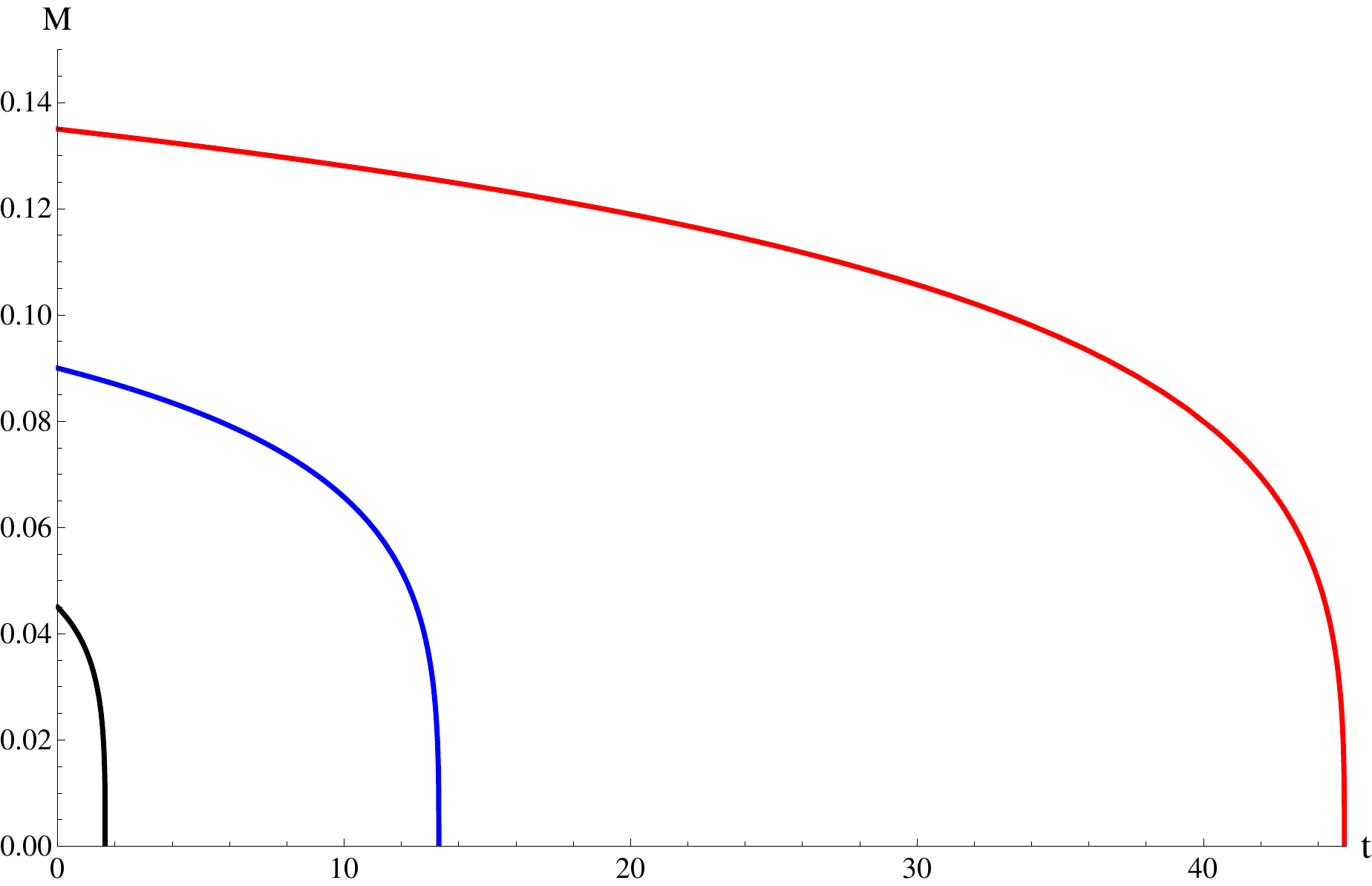}
\vspace{-1mm}
\caption{The evolution of the black hole for the case satisfying $-2\sqrt{\varepsilon+1}<\gamma\ell<-{3\sqrt{6(\varepsilon+1)}}/{4}$, $\varepsilon+1>0$, $\gamma<0$. We set $\varepsilon=0$, $\gamma\ell=-1.9$. For the left figure we consider arbitrarily large black hole evaporation (from $M\rightarrow\infty $ down to point \textsf{b} in FIG.\ref{fig9}). For the right figure we consider the black hole evaporation from point \textsf{d} to $M=0$. In each figure from left to right the curves correspond to $\ell=1$, $\ell=2$ and $\ell=3$ respectively.}
\label{fig11}
\end{center}
\end{figure}

In FIG.\ref{fig11} we present some examples of the black hole evaporation process for the two branches for fixed $y=-1.9$, $\varepsilon=0$, and $\ell=1,2,3$. For the first branch the lifetime is always infinite, while for the second branch the lifetime is of the order $\ell^3$.

\subsubsection{$0<M_c<M(r_1)<M(r_2)$}

\begin{figure}[h!]
\begin{center}
\includegraphics[width=0.45\textwidth]{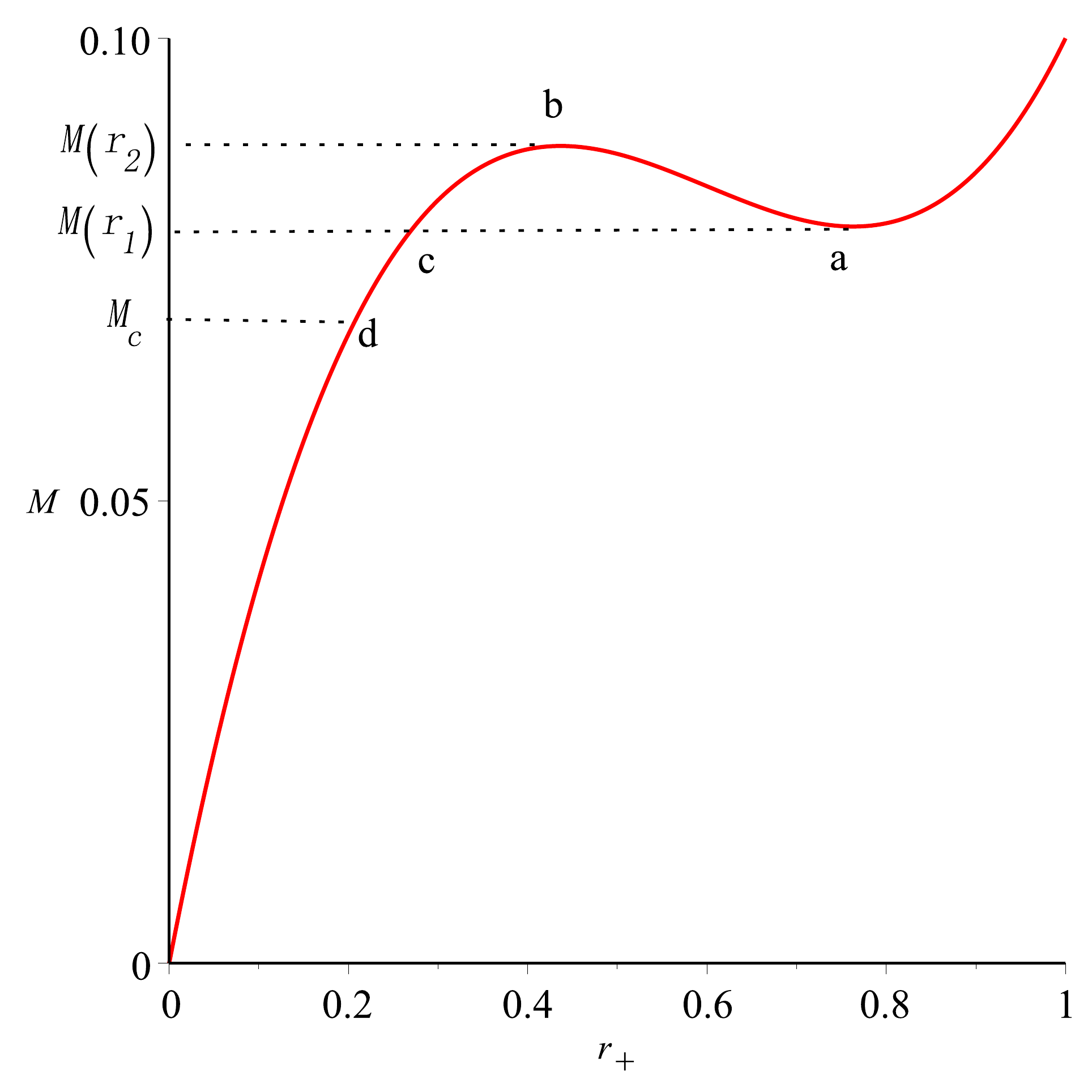}
\vspace{-1mm}
\caption{Behavior of the black hole mass $M$ as function of the horizon radius $r_+$ in $-{3\sqrt{6(\varepsilon+1)}}/{4}<\gamma\ell<-\sqrt{3(\varepsilon+1)}$, $\varepsilon+1>0$, $\gamma<0$. We set $\varepsilon=0$, $\gamma=-1.8$, and $\ell=1$. We have $0<M_c<M(r_1)<M(r_2)$.}
\label{fig12}
\end{center}
\end{figure}

Secondly we consider the case of $0<M_c<M(r_1)<M(r_2)$. This requires
\begin{equation}
-\frac{3\sqrt{6(\varepsilon+1)}}{4}<\gamma\ell<-\sqrt{3(\varepsilon+1)}.
\end{equation}
In FIG.\ref{fig12} we present the behavior of $M$ as the function of $r_+$. We set $\varepsilon=0$, $\gamma=-1.8$, $\ell=1$ so we have $0<M_c<M(r_1)<M(r_2)$. Similarly the solution is also branched and the region \textsf{a-b-c} should be excluded. For the branch from $M\rightarrow \infty$ to point \textsf{a}, the impact factor is $b_c=\ell$. For the branch from point \textsf{c} to $M=0$, since the black hole only admits one horizon, the impact factor is $b_c=\frac{r_{p1}}{\sqrt{f(r_{p1})}}$ as $M<M_c$, while $b_c=\ell$ as $M_c<M<M(r_1)$.

\begin{figure}[h!]
\begin{center}
\includegraphics[width=0.45\textwidth]{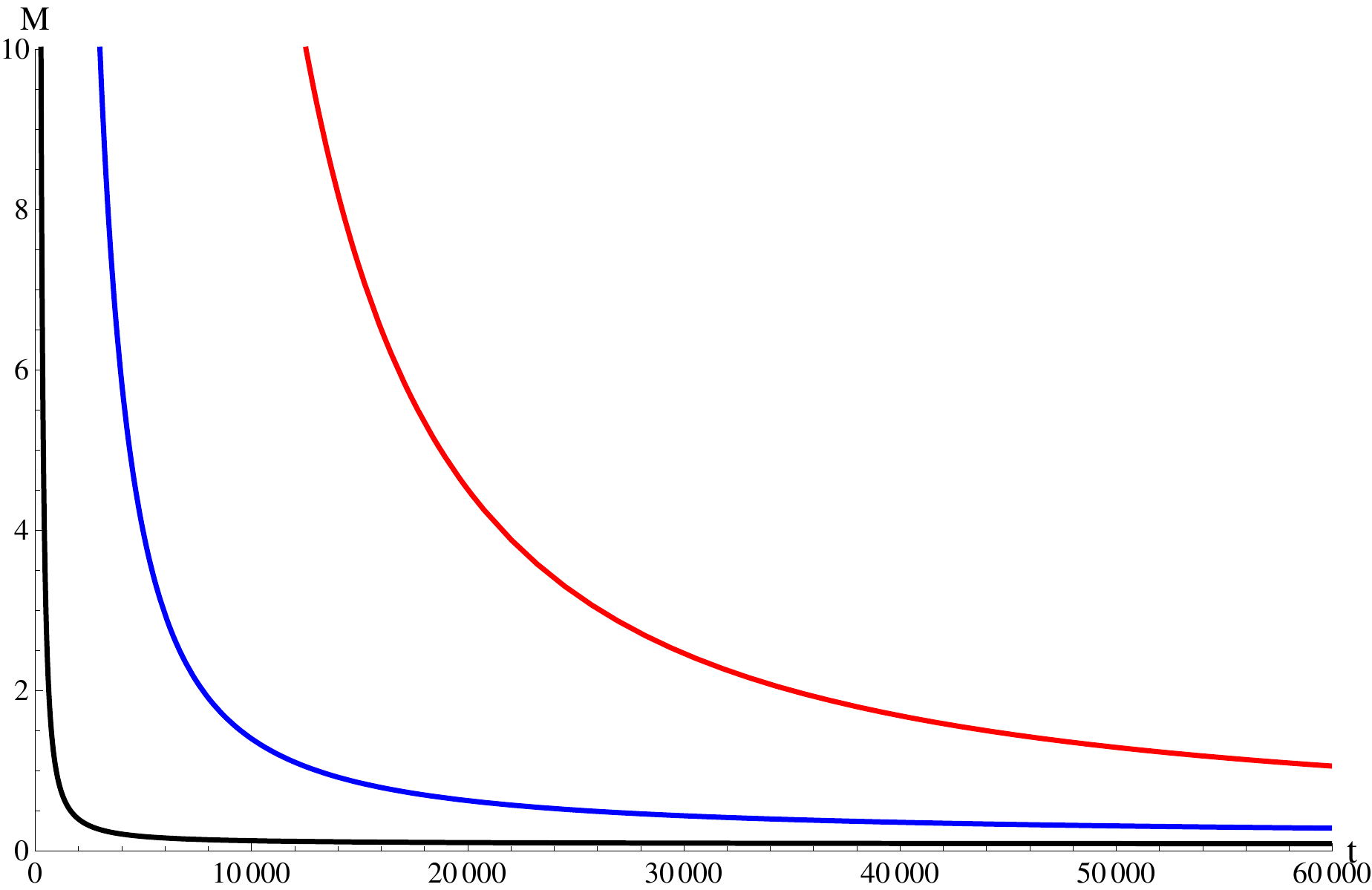}
\includegraphics[width=0.45\textwidth]{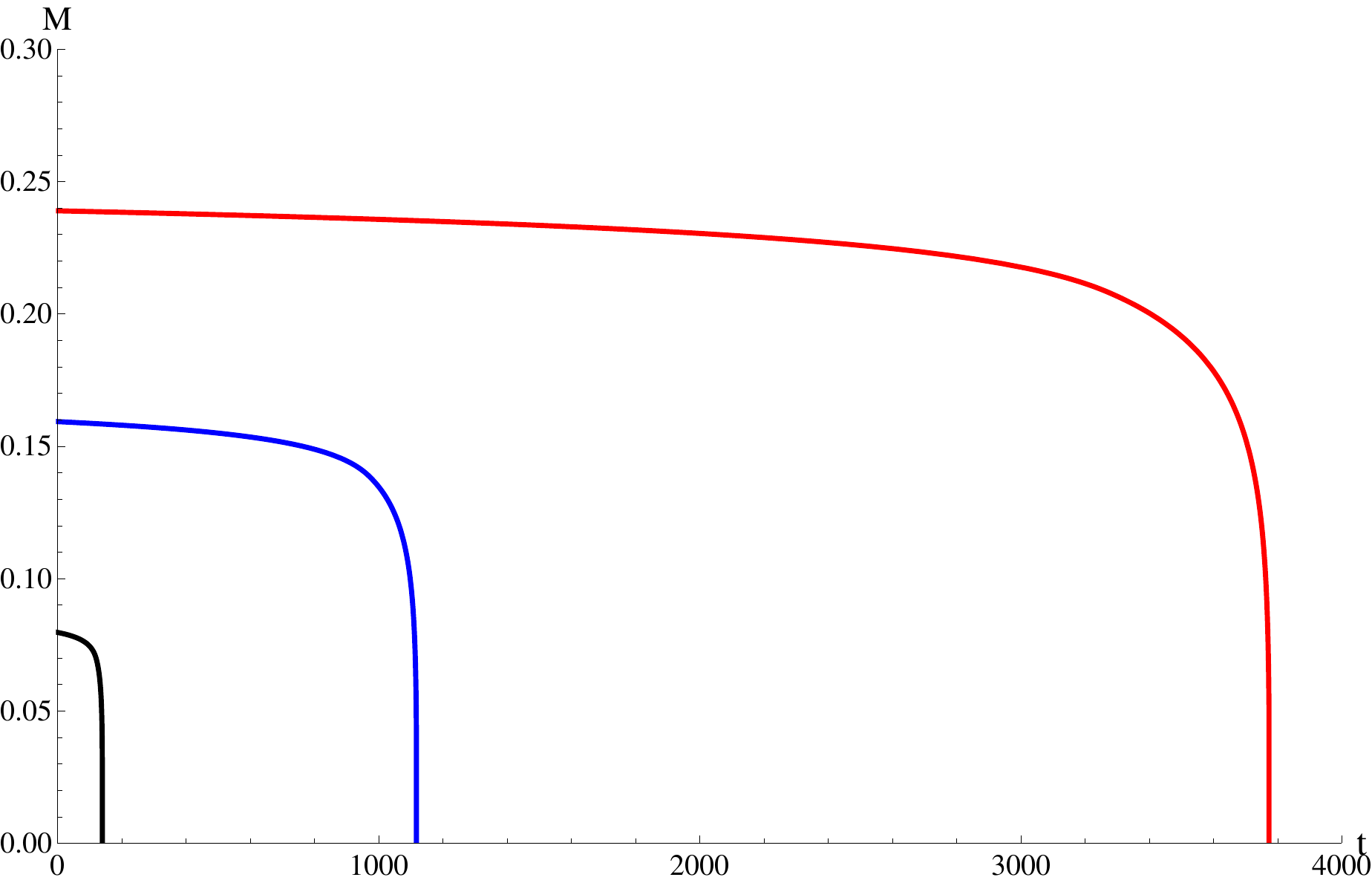}
\vspace{-1mm}
\caption{The evolution of the black hole for the case of $-\frac{3\sqrt{6(\varepsilon+1)}}{4}<\gamma\ell<-\sqrt{3(\varepsilon+1)}$, $\varepsilon+1>0$, $\gamma<0$. We set $\varepsilon=0$, $\gamma\ell=-1.8$. For the left figure we consider the arbitrarily large black hole evaporation ($M\rightarrow\infty $ to point \textsf{a} in FIG.\ref{fig12}). For the right figure we consider the black hole evaporation from point \textsf{c} to $M=0$. In each figure from left to right the curves correspond to $\ell=1$, $\ell=2$ and $\ell=3$, respectively.}
\label{fig13}
\end{center}
\end{figure}

In FIG.\ref{fig13} we present some examples of the black hole evaporation process for the two branches for fixed $y=-1.8$, $\varepsilon=0$, and $\ell=1,2,3$. For the branch from $M\rightarrow \infty$ to point \textsf{a}, since the temperature at point \textsf{a} is always zero, the black holes have infinite lifetime. For the branch from point \textsf{c} to $M=0$, we have
\begin{equation}
t=-\int^{x(c)}_{x_c}\frac{\partial \mathcal{M}(x,y,\varepsilon)}{\partial x}\frac{\ell^3}{\mathcal{T}^4(x,y,\varepsilon)}\d x-\int^{x_c}_{0}\frac{\partial \mathcal{M}(x,y,\varepsilon)}{\partial x}\frac{\ell^3}{\mathcal{B}^2(x,y,\varepsilon)\mathcal{T}^4(x,y,\varepsilon)}\d x,
\end{equation}
where $x(c)$ and $x_c$ denote the value of $x\equiv {r_+}/{\ell}$ at point \textsf{c} and \textsf{d} respectively. For fixed $y\equiv\gamma \ell$ and $\varepsilon$, both $x(c)$ and $x_c$ are constants, thus the integration on $x$ is also a constant. The lifetime is in order of $\ell$. In FIG.\ref{fig13} we present some numerical examples on both branches.

\subsubsection{$M(r_1)<0<M_c<M(r_2)$}

\begin{figure}[h!]
\begin{center}
\includegraphics[width=0.45\textwidth]{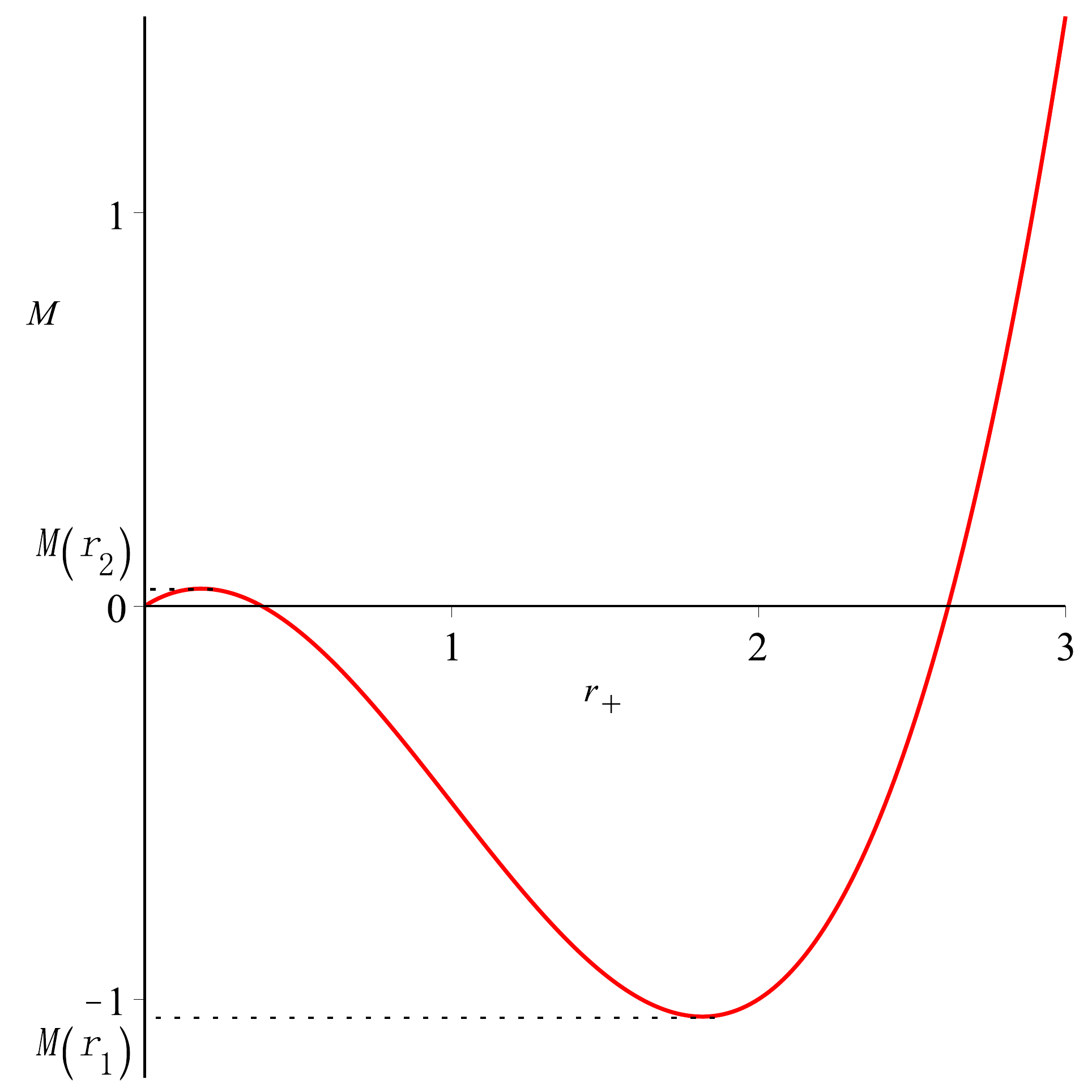}
\vspace{-1mm}
\caption{Behavior of the black hole mass $M$ as function of the horizon radius $r_+$ in $\gamma \ell<-2\sqrt{\varepsilon+1}$. We set $\varepsilon=0$, $\gamma=-3$, and $\ell=1$. We have $M(r_1)<0<M_c<M(r_2)$.}
\label{fig14}
\end{center}
\end{figure}

Lastly, we consider the case of $M(r_1)<0<M_c<M(r_2)$. The critical mass $M_c$ is always positive and smaller than $M(r_2)$, so we only need to consider the condition $M(r_1)<0$, which is $\gamma \ell<-2\sqrt{\varepsilon+1}$. In FIG.\ref{fig14} we present the example of $M$ as function of $r_+$. We set $\gamma=-3$, $\varepsilon=0$, $\ell=1$. Since the black hole radius is defined as the largest root, we know in this case the black hole has only one branch, which is from $M\rightarrow \infty$ to $M(r_1)$. The critical mass $M_c$ is located between $M=0$ and $M(r_2)$. As we have explained before, for cases with three roots, the maximal value of $\frac{f(r)}{r^2}$ is inside of the horizon, thus it is not part of the effective potential and the impact factor is still $b_c=\ell$.

\begin{figure}[h!]
\begin{center}
\includegraphics[width=0.45\textwidth]{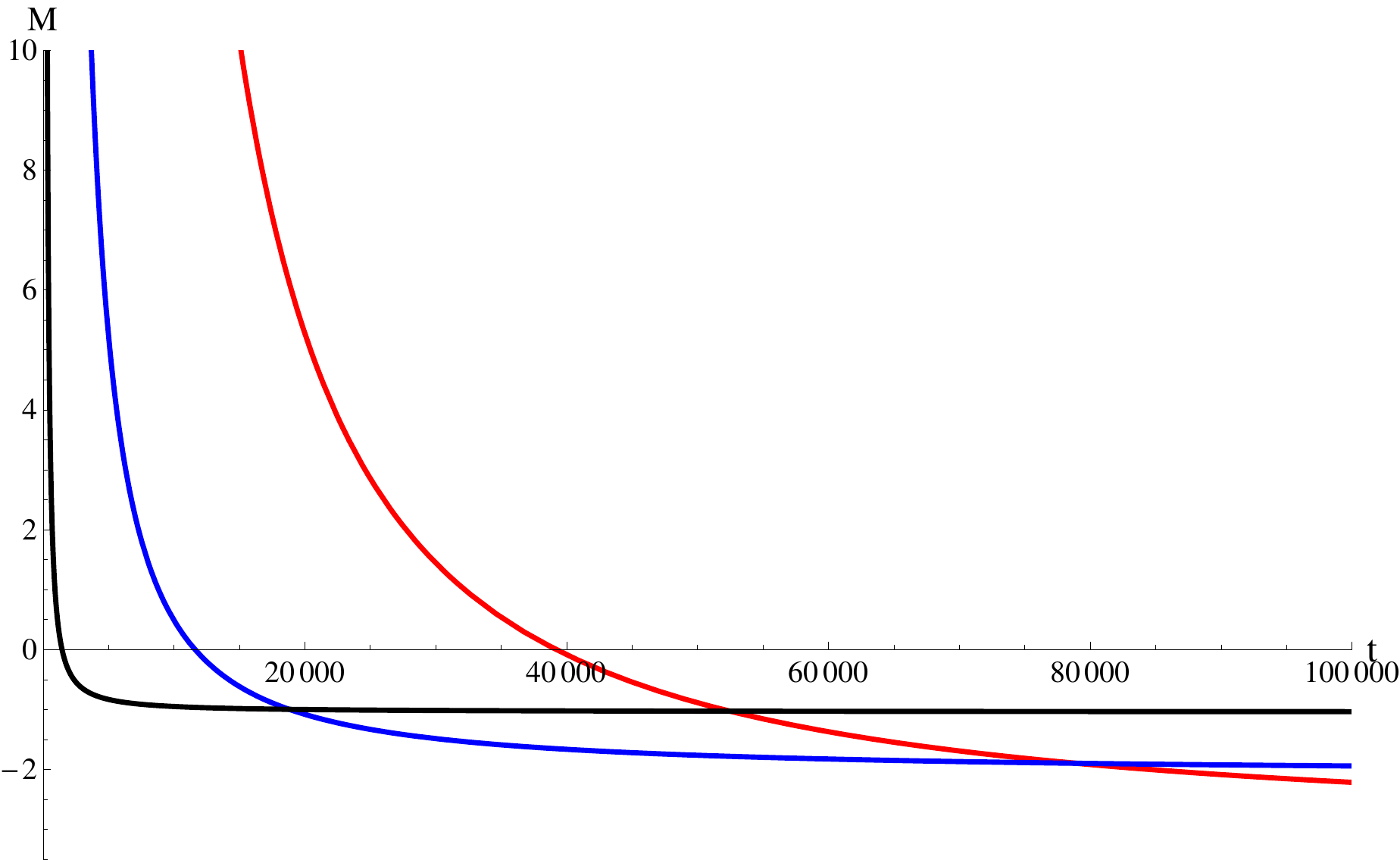}
\vspace{-1mm}
\caption{The evolution of the black hole for the case of $\gamma \ell<-2\sqrt{\varepsilon+1}$, $\varepsilon+1>0$, $\gamma<0$. We set $\varepsilon=0$, $\gamma \ell=-3$. From left to right the curves correspond to $\ell=1$, $\ell=2$ and $\ell=3$, respectively.}
\label{fig15}
\end{center}
\end{figure}

In FIG.\ref{fig15} we present some examples of black hole evaporation in $\varepsilon=0$, $\gamma \ell=-3$ and various $\ell$. There is only one branch and the final state $M(r_1)$ corresponds to $T=0$, so the black hole lifetime is always infinite, satisfying the third law of black hole thermodynamics.

\section{Conclusion}

Massive gravity is an extension of general relativity by a non-zero graviton mass term. For years it suffers from the vDVZ discontinuity and BD ghost, but a new model, known as the dRGT massive gravity, revived the interests in massive gravity. By choosing the coefficients of the effective field theory order by order, dRGT massive gravity leads to the resummation of the entire infinite series of the terms in the effective Lagrangian, and the field equation is at most second order in time derivatives. Such ghost-free theory still appears to be somewhat problematic, however it is useful as an effective field theory in holographic applications to allow momentum dissipation in the dual field theory. In view of holography, black hole solutions in dRGT gravity deserve further investigation.

In the present work we investigate a class of $(3+1)$-dimensional spherically symmetric evaporating black holes of dRGT massive gravity in AdS spacetime. The graviton mass term generates three terms in the black hole metric, which are, an effective cosmological constant term ${r^2}/{\ell^2}$, a linear term $\gamma r$, and a global monopole term $\varepsilon$. These terms modify the thermodynamical properties of general relativity black hole. Unlike the well-known Schwarzschild-AdS case, the black hole thermodynamics in dRGT massive gravity is quite rich. There are cases with more than one horizon, as well as existence of effective black hole remnant. Applying the geometrical optics approximation and an absorbing AdS boundary condition, we can study the black hole evaporation by Stefan-Boltzmann law \eqref{law}. The effective emission surface can be proportional to the square of the effective AdS length, or the square of the impact parameter corresponding to the photon orbit. For certain cases it is also possible that one emission surface changes to another one as the black hole losses its mass. Since the $T^4$ term is of a higher order, the behavior of the temperature $T$, especially the asymptotic behavior, is very important in the process of black hole evaporation.

\begin{figure}[h!]
\begin{center}
\includegraphics[width=0.45\textwidth]{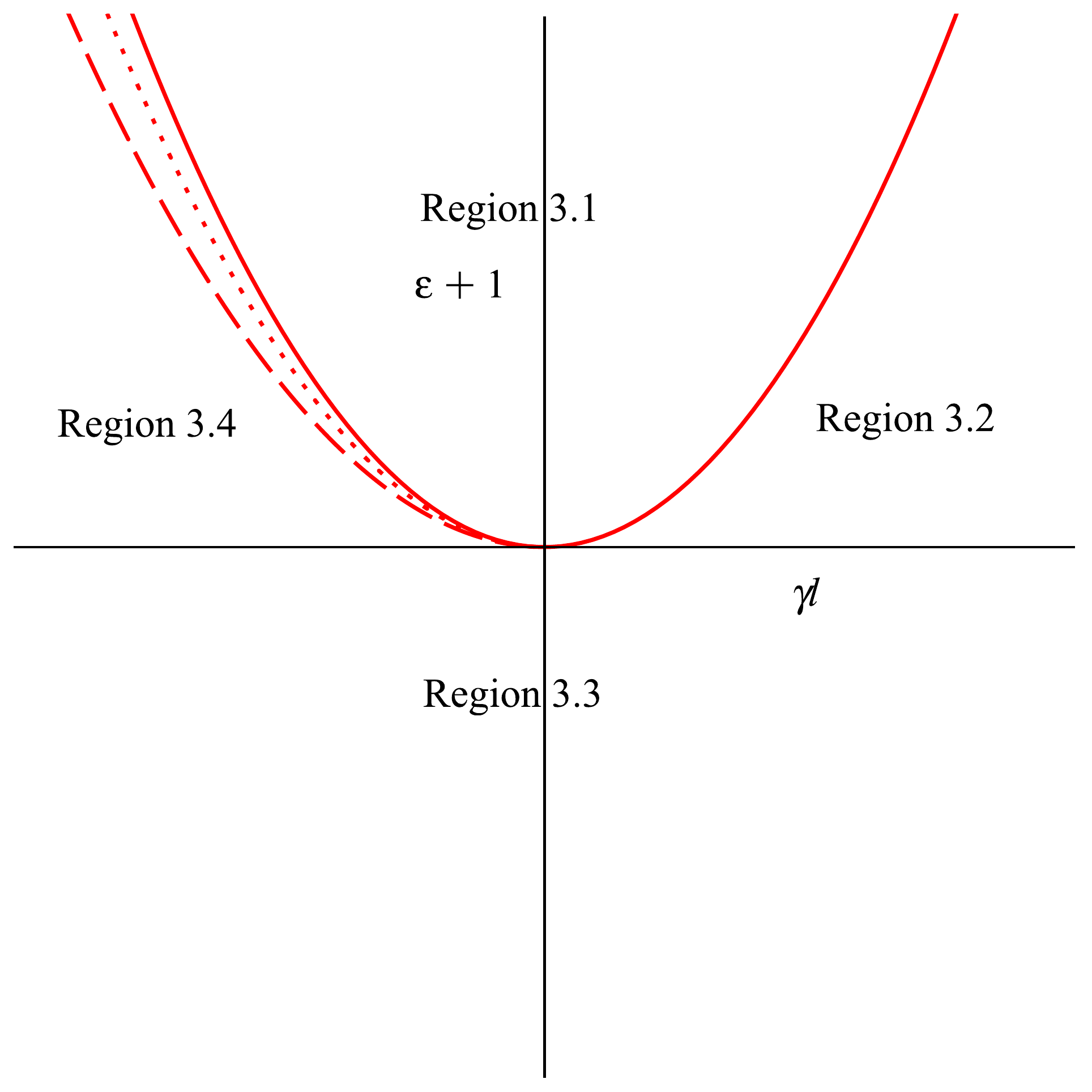}
\vspace{-1mm}
\caption{The parameter region in ($\gamma \ell,\varepsilon+1$) plane. The solid curve $\gamma^2\ell^2=3(\varepsilon+1)$ and the axis $\gamma \ell$ separate the region into four parts, which correspond the cases in Sec.3.1, 3.2, 3.3, 3.4. The dotted and dashed curves, respectively, are $\gamma\ell=-\frac{3\sqrt{6(\varepsilon+1)}}{4}$ and $\gamma\ell=-2\sqrt{\varepsilon+1}$ that separate Region 3.4 into three parts that correspond to the different features of $M$ in Sec.3.4. Specifically, the portion between horizontal axis and the dashed curve corresponds to Sec.3.4.3, the portion between the dashed curve and the dotted curve corresponds to Sec.3.4.1, and the portion between the dotted curve and the solid curve is discussed in Sec.3.4.2.}
\label{fig16}
\end{center}
\end{figure}

We consider different cases of black hole evaporation, which we now summarized. Depending on the features of zero points of $T$, we can divide the parameter region into four parts in ($\gamma \ell,\varepsilon+1$) plane by the solid line $\gamma^2\ell^2=3(\varepsilon+1)$ and the axis $\gamma \ell$ in FIG.\ref{fig16}. When the zero points of $T$ are both unreal or negative (Region 3.1 and 3.2), setting $\gamma\ell=\text{const}$, the lifetime of arbitrarily large black hole is in the order of $\ell^3$. When the temperature admits only one positive root (Region 3.3), the lifetime of the black hole is infinite, and the black hole effectively becomes an effective remnant near the $T=0$ state, which is in accordance to the third law of black hole thermodynamics. For the cases of two positive zero points $T$ (Region 3.4), the black hole evaporation process also depends on the features of $M$. The black hole solution can be branched (region between the dashed line $\gamma\ell=-2\sqrt{\varepsilon+1}$ and solid line $\gamma\ell=-\sqrt{3(\varepsilon+1)}$), with one branch admitting infinite lifetime while the other is of order $\ell^3$. Alternatively, there could also be only one single branch (region between the dashed line $\gamma\ell=-2\sqrt{\varepsilon+1}$ and axis $\gamma \ell$) and the black hole lifetime is always infinite in that case.

Of course we can also consider the cases of $\varepsilon+1=0$ and $\gamma^2\ell^2=3(\varepsilon+1)$. For $\varepsilon+1=0$ (that is, on the horizontal axis of FIG.(\ref{fig16})), we have the temperature
\begin{equation}
T=\frac{1}{4\pi}\left(\frac{3r_+}{\ell^2}+2\gamma \right),
\end{equation}
which admits one zero point at $r_+=-\frac{2}{3}\gamma \ell^2$. For $\gamma\leqslant 0$ the final state corresponds to a remnant, thus the lifetime of black holes is infinite.
For $\gamma>0$ the lifetime is also in the order of $\ell^3$ as $\gamma\ell=$const. Note that for $\ell \to \infty$ we have a peculiar feature in which the black hole has a \emph{constant} temperature. The lifetime of this black hole is infinite as shown in \cite{1812.03136} -- it is an example of the ``complementary third law'', in which under some reasonable assumptions, it was proved therein that if towards the end the temperature is finite and nonzero yet the black hole goes to zero size, then such a state is unattainable in finite time. This is consistent with the result here since this amounts to a lifetime of $\ell^3$, which of course tends to infinity.

In the case of $\gamma^2\ell^2=3(\varepsilon+1)$, we have $r_1=r_2=-\frac{\gamma\ell^2}{3}$. For $\gamma>0$ the lifetime is of the order $\ell^3$, while for $\gamma<0$ the solution is branched and we have the point \textsf{a} and \textsf{b} coincide in FIG.\ref{fig12}. One branch admits infinite lifetime while the other is of order $\ell^3$. Similarly if we consider the cases of $\gamma\ell=-{3\sqrt{6(\varepsilon+1)}}/{4}$ and $\gamma\ell=-2\sqrt{\varepsilon+1}$, we will find that they correspond to $M_c=M(r_1)$ and $M(r_1)=0$, respectively. The qualitative features of the black hole evaporation stay the same.

This also means that the discussion in \cite{EslamPanah:2018rob} is too simplistic. In that work, it was argued that if our universe is fundamentally anti-de Sitter-like with a transient accelerating phase, then massive gravity could result in black hole remnants that could in turn ameliorate the information paradox (and possibly also provide an explanation for dark matter). From the discussions in this work, we now know that the various parameters have to be chosen with care in order for this to happen.

\section*{Acknowledgements}

YCO thanks the National Natural Science Foundation of China (No.11705162, No.11922508) and the Natural Science Foundation of Jiangsu Province (No.BK20170479) for funding support.


\providecommand{\href}[2]{#2}\begingroup
\footnotesize\itemsep=0pt
\providecommand{\eprint}[2][]{\href{http://arxiv.org/abs/#2}{arXiv:#2}}

\end{document}